\documentclass[a4paper,10pt]{article}

\usepackage[a4paper,left=3cm,right=3cm,top=2.5cm,bottom=2.5cm]{geometry}
\usepackage{hyperref}
\usepackage{subfig}
\usepackage{pgfplots}
\usepackage{pgfplotstable}
\usepackage{mathtools}
\usepackage{multicol}
\usepackage{comment}
\usepackage{booktabs}
\pgfplotsset{compat=1.5}
\usepackage{amssymb}
\usepackage{url}
\usepackage{bm}
\usepackage{tikz} 
\usetikzlibrary{shapes,arrows,positioning,chains,fit,arrows.meta}
\usepackage{textcomp}

\usepackage{float}
\usepackage{tabularx}
\usepackage{enumerate}
\usepackage{array}
\usepackage{xspace}
\usepackage{siunitx}
\usepackage{authblk}

\usepackage{amsthm}
\usepackage{amsmath}
\usepackage{stmaryrd}
\usepackage[ruled,vlined,linesnumbered]{algorithm2e}
\usepackage{graphicx}
\usepackage{booktabs}
\usepackage{cleveref}

\DeclareFontEncoding{LS1}{}{}
\DeclareFontSubstitution{LS1}{stix}{m}{n}
\DeclareSymbolFont{symbolsstix}{LS1}{stixscr}{m}{n}
\DeclareMathSymbol{\M}{0}{symbolsstix}{'115}
\DeclareMathSymbol{\SSS}{0}{symbolsstix}{'123}
\DeclareMathSymbol{\s}{0}{symbolsstix}{'163}
\DeclareMathSymbol{\R}{0}{symbolsstix}{'122}
\DeclareMathSymbol{\m}{0}{symbolsstix}{'155}
\DeclareMathSymbol{\n}{0}{symbolsstix}{'156}
\DeclareMathSymbol{\kk}{0}{symbolsstix}{'153}

\newcommand{\fig}{\text{Fig.~}}
\newcommand{\tab}{\text{Tab.~}}
\newcommand{\eq}{\text{Eq.~}}
\newcommand{\sez}{\text{Sec.~}}

\begin{document}

\title{Real-time topology optimization via learnable mappings}

\author{Gabriel~Garayalde\footnote{gabrielemilio.garayalde@mail.polimi.it}}
\author{Matteo~Torzoni\footnote{matteo.torzoni@polimi.it}}
\author{Matteo~Bruggi\footnote{matteo.bruggi@polimi.it}}
\author{Alberto~Corigliano\footnote{alberto.corigliano@polimi.it}}

\affil{Dipartimento di Ingegneria Civile e Ambientale, Politecnico di Milano,\\Piazza L. da Vinci 32, 20133 Milan, Italy}

\maketitle

\begin{abstract}

In traditional topology optimization, the computing time required to iteratively update the material distribution within a design domain strongly depends on the complexity or size of the problem, limiting its application in real engineering contexts. This work proposes a multi-stage machine learning strategy that aims to predict an optimal topology and the related stress fields of interest, either in 2D or 3D, without resorting to any iterative analysis and design process. The overall topology optimization is treated as regression task in a low-dimensional latent space, that encodes the variability of the target designs. First, a fully-connected model is employed to surrogate the functional link between the parametric input space characterizing the design problem and the latent space representation of the corresponding optimal topology. The decoder branch of an autoencoder is then exploited to reconstruct the desired optimal topology from its latent representation. The deep learning models are trained on a dataset generated through a standard method of topology optimization implementing the solid isotropic material with penalization, for varying boundary and loading conditions. The underlying hypothesis behind the proposed strategy is that optimal topologies share enough common patterns to be compressed into small latent space representations without significant information loss. Results relevant to a 2D Messerschmitt-Bölkow-Blohm beam and a 3D bridge case demonstrate the capabilities of the proposed framework to provide accurate optimal topology predictions in a fraction of a second.

\end{abstract}

\section{Introduction}
Topology optimization (TO) is a robust tool for computational design, capable of creating intricate and high-performance structures that adhere to a set of prescribed design constraints, in terms of loads and boundary conditions. TO methods optimize the distribution of available material within a specified domain by finding layouts that minimize or maximize a suitable objective function, such as minimizing compliance or maximizing thermal conductivity \cite{Bendsoee1999, Bejan1997, art:Weihong}. The resulting topology provides an efficient use of material within the design domain, yielding cost savings and lightweight structures -- all valuable features for engineers.

The most widely used numerical TO methods rely on the so-called solid isotropic microstructure with penalty (SIMP) material model. Proposed in the landmark paper by Bendsoe (1989) \cite{Bendsoee1989}, it has rapidly gained momentum to become the subject of many research activities in the following decades \cite{Rozvany1992, Zhou1991, Rozvany2000, Bendsoee2003}. Within a SIMP-based TO approach, an iterative scheme is needed to update the design variable, i.e. the point-wise material density. At each step, it is necessary to evaluate the sensitivity of the objective function and the constraints with respect to the minimization unknown. This calls for solving the state equation iteratively. Despite the increasing popularity of TO, its effective application in large-scale engineering contexts and real-time applications is limited by the associated computational burden. To address this challenge, researchers are exploring, among other approaches, the use of artificial intelligence to engineer novel solutions capable of reducing computational times and accelerating the TO process.

In 2017, Sosnoviks and Oseledets \cite{Sosnovik2017} proposed one of the first deep learning (DL) approaches to two-dimensional (2D) TO, framed as an image segmentation task addressed with a deep convolutional neural network (CNN). The final black-and-white topology is predicted by providing as input to the neural network two grayscale images: the density distribution and the gradient of the densities, both attained after only a handful of iterations of the SIMP solver. This method is coined as an \textit{accelerative} method, see e.g.~\cite{Lin2018, Kallioras2020, Banga2018, Senhora2022, Ma2023}. Indeed, the TO algorithm is exploited only for a limited number of iterations, whilst the more time consuming refinement phase is replaced with the DL model, greatly speeding up the procedure.

\textit{Non-iterative} optimization methods, unlike accelerative ones, exploit the use of artificial intelligence algorithms to predict an optimal shape that conforms to the design constraints, without an iterative form finding process. These have been applied for predicting three-dimensional (3D) topologies, yielding good generalization capabilities for variable design domains and different loading and boundary conditions, see e.g.~\cite{DiabAbueidda2020, Kollmann2020, Zhang2021, Rawat2018, Rawat2019}. For instance, in \cite{Zheng2021}, the boundary conditions, loading conditions, volume fraction and domain size are directly encoded into the input tensor of a U-Net autoencoder (AE) architecture that is exploited to predict the corresponding optimal topology. An alternative solution was proposed in \cite{Herath2021}, through the combined use of CNNs and conditional generative adversarial networks, which can generate the optimal design and trace the corresponding Von Mises (VM) stress contour, eventually allowing the identification of the most stressed regions.

A subset of non-iterative methods is that of \textit{non-iterative physics-aided} methods, which exploit the physics behind the problem in addition to the data -- either incorporated as an additional term in the training loss function to be minimized or in a principled information encoding in the input and target data tensors, see e.g.~\cite{Cang2018, Xiang2022a, Jeong2023, Yan2022, Finol2019, Asad2022, Odot2022}. Although some of these methods require a first run of the finite element (FE) algorithm to generate an initial field of interest, for instance in terms of displacements or strains, they are still considered as non-iterative methods. Relevant recent contributions in non-iterative physics-aided methods include: the U-net-based strategy proposed in \cite{Zhang2019}, which exploits an input tensor containing information about the prescribed volume fraction, and the displacement and strain fields, as obtained from a SIMP-based procedure; its extension to 3D and stress minimization problems proposed in \cite{Xiang2022}; the ``TopologyGAN'' framework \cite{Nie2021}, which exploits the strain energy density and the VM stress field obtained from the FE analysis, along with the prescribed volume fraction, boundary and loading conditions, as the inputs to the generator of a conditional generative adversarial network; the ``GANTL'' framework \cite{Behzadi2022}, which incorporates similar input fields to the generative adversarial network as in \cite{Nie2021}, but also exploits a multi-fidelity transfer learning strategy to retrain a source network, previously trained on a large set of low-resolution images, on high-resolution images obtained for varying boundary conditions.

This paper proposes a multi-stage DL model, capable of predicting an optimal topology in a single shot, either in 2D or 3D, by exploiting a DL-based dimensionality reduction of the target design. To obtain a low-dimensional non-linear projection of the TO solution, the proposed method employs a strategy similar to the reduced-order modeling approach for parametrized PDEs proposed in \cite{Fresca2020}. In particular, a fully-connected model is employed to map the parametric input space characterizing the design problem onto the latent representation of the corresponding optimal topology, and the decoder branch of an AE model is then exploited to reconstruct the desired optimal topology from its latent representation. 

The advantage of the proposed method stems from the capability of exploiting common patterns that characterize the topologies in the training dataset, thus enabling an efficient data compression without significant information loss. This allows learning an effective functional link between the parametric input space and the low-dimensional latent space through the fully-connected model. The input parameters to the problem are encoded efficiently as a compact numerical array, instead of highly sparse images, see e.g.~\cite{DiabAbueidda2020, Xiang2022a}. Furthermore, our method is purely data-driven, eliminating the need for physics-based solvers during the prediction phase. This stands in contrast to other approaches, e.g., \cite{Nie2021} and \cite{Behzadi2022}, which rely on mechanical information from physics-based solvers as input. The primary advantage lies in the increased computational efficiency, especially when scaling to larger, more complex problems, as there is no need to run potentially resource-demanding simulations during prediction. This efficiency stands as one of the key capabilities of purely data-driven methods, making our approach ideal for real-time applications and iterative design processes where rapid predictions are crucial. Additionally, our approach can be adapted to handle highly complex or non-linear problems, which may face computational limitations when relying on physics-based solvers for multi-query applications. On the other hand, it is important to note the risk of inconsistent solutions associated with the purely-data driven perspective, highlighting the need for careful consideration.

Our procedure, however, does utilize physics-based information in the training stage, namely the fields of VM stress and dominant principal stress, which are encoded along with the optimized density field. Once trained, the proposed framework has the advantage of not only predicting the optimal density field, similarly to \cite{Kallioras2020, DiabAbueidda2020, Nie2021, Behzadi2022}, but also the related stress fields, all almost instantaneously. Our framework thus offers clear advantages over traditional and computationally expensive TO algorithms. This can be especially useful in the conceptual design phase of a project, when it is necessary to investigate feasible solutions quickly and accurately. The proposed computational framework is offered with an online app to showcase the potentialities of real-time topology optimization, which is made available at \cite{app}.

The reminder of this paper is organized as follows. In \sez\ref{sec:methodology}, we describe the proposed methodology. In \sez\ref{sec:experiments}, the whole computational procedure is then assessed on two test cases, respectively related to a 2D Messerschmitt-Bölkow-Blohm (MBB) beam and a 3D bridge case. Conclusions and future perspectives are finally drawn in \sez\ref{sec:conclusions}.

\section{Methodology}
\label{sec:methodology}

In this section, we describe the methodology characterizing the proposed framework in terms of: the composition of the handled data in \sez\ref{sec:dataset_creation}; the physics-based numerical model and TO formulation behind the simulation-based generation of labeled training dataset in \sez\ref{sec:TO_formulation}; the setup of the ML strategy, employed to approximate optimal designs in real-time in \sez\ref{sec:ML_formulation}.

\subsection{Data specification}
\label{sec:dataset_creation}

For an assigned design domain discretized into $N_e$ finite elements, the training dataset $\boldsymbol{D}$ is built upon $ I $ instances provided through a FE-based TO algorithm, as follows:
\begin{equation}
\boldsymbol{D} = \{  (\boldsymbol{x}, \boldsymbol{x}_\mathtt{VM},\boldsymbol{x}_\mathtt{TC}, \boldsymbol{p})_i \}^{I}_{i=1}.
\label{eq:dataset}
\end{equation}
Each instance consists of the optimal topology $\boldsymbol{x}\in\mathbb{R}^{N_e}$, obtained for a given vector $\boldsymbol{p}\in\mathbb{R}^\mathtt{N_{par}}$ of $\mathtt{N_{par}}$ parameters, ruling the applied loading and boundary conditions, and of the corresponding fields of VM stress $\boldsymbol{x}_\mathtt{VM}\in\mathbb{R}^{N_e}$ and of dominant principal stress $\boldsymbol{x}_\mathtt{TC}\in\mathbb{R}^{N_e}$. After performing a suitable data normalization, the three fields can take values in the following ranges $\boldsymbol{x}\in[0,1]$, $\boldsymbol{x}_\mathtt{VM}\in [0,1]$ and  $\boldsymbol{x}_\mathtt{TC}\in [-1,1]$. To populate $\boldsymbol{D}$, the parametric input space described by $\boldsymbol{p}$ is assumed to be uniformly distributed, and then sampled via latin hypercube rule. It is worth noting that both the probability distribution modeling $\boldsymbol{p}$ and the sampling strategy used to explore it have an impact on the distribution of various optimal topologies collected in $\boldsymbol{D}$. This impact is contingent on the characteristics of the problem at hand. A sensitivity analysis could identify the ranges of input parameters where larger variations in achieved topologies are observed. Such an analysis could be beneficial for creating datasets that are more balanced in terms of topology varieties, or for generating smaller datasets that still adequately cover the space optimal designs. While we recognize the value of this type of investigation, it falls beyond the scope of this work. Therefore, the training and testing data in this work are generated by exploring the parametric input space underlying $\boldsymbol{p}$ as described above, although there are no restrictions in this respect. In the reminder of the paper, the index $i$ will be dropped for ease of notation, unless necessary.

\subsection{Topology optimization formulation}
\label{sec:TO_formulation}

With reference to compliance minimization problems in the presence of statically applied loads, the training dataset $\boldsymbol{D}$ is populated according to \eq\eqref{eq:dataset} by means of the SIMP approach as implemented in \cite{Sigmund2001} and \cite{Ferrari2020} for 2D and 3D problems, respectively. In particular, within a discrete setting relying on the FE method to solve the linear elastic equilibrium, each finite element of the FE mesh is characterized by a density value $x_{e}\in [0,1]$, where $e=1,\ldots,N_e$ and $N_e$ is the total number of elements. The corresponding value of the Young's modulus $E_e$, reads as:
\begin{equation}
    \label{eq:youngs_modulus_1}
    E_e(x_{e}) = E_{min}+ x_{e}^{p}(E_0 - E_{min}).
\end{equation}
Herein: $E_{min}$ is a threshold stiffness value to avoid singularities in the stiffness matrix; $E_0$ is the reference value of the material stiffness; the exponent $p$ is a penalization factor to ensure black-and-white solutions. Under the assumption of a linearized kinematics, the corresponding TO problem is formulated as:
\begin{equation}
\begin{split}
\min_{\boldsymbol{x}}: & \quad c(\boldsymbol{x}) = \boldsymbol{U}^T\boldsymbol{K}\boldsymbol{U} = \sum_{e=1}^{N_e} E_e(x_{e})\boldsymbol{u}_e^T\boldsymbol{k}_0\boldsymbol{u}_e, \\
\text{subject to}: & \quad \boldsymbol{K}\boldsymbol{U} = \boldsymbol{F}, \\
& \quad V(\boldsymbol{x})/V_0 = f, \\
& \quad \boldsymbol{0} \leq \boldsymbol{x} \leq \boldsymbol{1},
\end{split}
\label{eq:opt}
\end{equation}
with $c(\boldsymbol{x})\in\mathbb{R}^{N_e}$ being the structural compliance to be minimized with respect to the design variables $\boldsymbol{x}$. In \eq\eqref{eq:opt}: $\textbf{U}\in\mathbb{R}^N$ is the global vector of nodal displacements; $N$ is the number of degrees of freedom; $\textbf{K}\in\mathbb{R}^{N\times N}$ is the global stiffness matrix, assembled under the assumption of a linear elastic isotropic material; $\textbf{u}_e$ is the local displacement vector for the $e$-th element; $\textbf{k}_0$ is the element stiffness matrix for an element featuring a ``full material'' Young's modulus; $\textbf{F}\in\mathbb{R}^N$ is the global vector of nodal forces; $f\in\mathbb{R}$ is the prescribed volume fraction; $V(\boldsymbol{x})\in\mathbb{R}$ and $V_0\in\mathbb{R}$ refer to the volume of the placed material and to the volume of the design domain, respectively. 

At each iteration of the optimization procedure, the elements densities $\boldsymbol{x}$ are updated employing a standard optimality criterion method, based on the following heuristic updating strategy:
\begin{equation}
\label{eq:optimality_criteria_method}
    x_{e}^{new} = 
    \left\{
    \begin{array}{ll}
    \text{max}(0, x_{e} - m) & \text{if} \,\, x_e B_e^\eta \, \leq \, \text{max}(0, x_{e} - m), \\
    \text{min}(1, x_{e} + m) & \text{if} \,\, x_{e} B_e^\eta \, \geq \, \text{min}(1, x_{e} + m), \\
    x_{e} B_e^\eta &\text{otherwise},
    \end{array}
    \right.
\end{equation}
where $m\in\mathbb{R}$ is a suitable clipping value on the iteration step, $\eta = \frac{1}{2}$ is a numerical damping coefficient, and $B_e$ is obtained as:
\begin{equation}
    \mathbf{B}_e=\frac{-\frac{\partial c}{\partial x_{e}}}{\lambda\frac{\partial V}{\partial x_{e}}},
    \label{eq:Be}
\end{equation}
with $\lambda\in\mathbb{R}$ being a suitable Lagrangian multiplier computed through a bisection algorithm. The sensitivities with respect to the element densities $x_e$ involved in \eq\eqref{eq:Be}, namely the sensitivity of $c$ and of $V$, are computed under the assumption that each element features a unit volume, yielding:
\begin{equation}
    \begin{split}
    \frac{\partial c}{\partial x_{e}} = & -px_{e}^{p-1} (E_0 - E_{min}) \boldsymbol{u}_e^T\boldsymbol{k}_0\boldsymbol{u}_e,\\
    \frac{\partial V}{\partial x_{e}} = & 1.
    \end{split}
    \label{eq:sensitivity}
\end{equation}

To avoid undesired checkerboard patterns (unphysical minima) and mesh dependence issues, the filtering strategy proposed in \cite{Sigmund1994} is used to compute a filtered sensitivity $\widehat{\frac{\partial c}{\partial x_{e}}}$, as follows:
\begin{equation}
\widehat{\frac{\partial c}{\partial x_{e}}} = \frac{1}{\text{max}(\gamma,x_{e})\sum_{j \in N_e}H_{ej}} \sum_{j \in N_e}H_{ej}x_{ej}\frac{\partial c}{\partial x_{e}},
    \label{eq:filtering}
\end{equation}
which involves a weighted average over different elements, that regularizes the space of solutions to the optimization problem. In \eq\eqref{eq:filtering}: $H_{ej}= \text{max}(0,r_{min}-\Delta(e,j))$ is the weight factor for the $j$-th set of elements, for which the center-to-center distance to the $e$-th element is smaller than the filter radius $r_{min}$; $\gamma =10^{-3}$ is a small positive term to avoid zero division. This differs from the classical SIMP approach \cite{Sigmund2001}, in which the density variables $x_e$ can not become zero, thus not requiring $\gamma$ to be defined. It has to be remarked that the use of a filtering scheme may imply the presence of gray boundaries in the resulting optimal design. To get a crispier transition between dense and void regions, projection schemes may be conveniently implemented, as discussed in \cite{Guest2004}. For a more detailed analysis of filtering methods, the reader is referred to \cite{Sigmund2007}. 

While the case of TO 2D problems is based on the implementation code \cite{Sigmund2001} following the above formulation, the case of 3D TO problems is instead addressed following the ``top3D125'' implementation \cite{Ferrari2020}, which presents some minor modifications for the filtering strategy. In particular, a density filter is exploited to transform the original design variables $x_e$ into the so called physical densities $\widetilde{x_{e}}$, as follows:
\begin{equation}
     \widetilde{x_{e}} = \frac{1}{\sum_{j \in N_e}H_{ej}} \sum_{j \in N_e}H_{ej}x_{j}.
    \label{eq:filtering_densities}
\end{equation}

 The use of the above filtered densities requires a minor reformulation of the sensitivities computation. Indeed, the sensitivities of the objective function $c$ and of the material volume $V$ with respect to the physical densities $\widetilde{x_{e}}$ are still given by \eq\eqref{eq:sensitivity}, provided that $x_{e}$ is replaced with $\widetilde{x_{e}}$ by means of the chain rule of derivation, as follows: 
\begin{equation}
\begin{split}
    \frac{\partial c}{\partial x_{k}} = 
    \sum_{e \in N_k} \frac{\partial c}{\partial \widetilde{x_{e}}} \frac{\partial \widetilde{x_{e}}}{\partial x_{k}} =
    \sum_{e \in N_k} \frac{1}{\sum_{j \in N_e}H_{ej}} H_{ke} \frac{\partial c}{\partial \widetilde{x_{e}}},\\
    \frac{\partial V}{\partial x_{k}} = 
    \sum_{e \in N_k} \frac{\partial V}{\partial \widetilde{x_{e}}} \frac{\partial \widetilde{x_{e}}}{\partial x_{k}} =
    \sum_{e \in N_k} \frac{1}{\sum_{j \in N_e}H_{ej}} H_{ke} \frac{\partial V}{\partial \widetilde{x_{e}}}.
\end{split}
\label{eq:chain_rule}
\end{equation}

At the end of each topology optimization loop while populating $\boldsymbol{D}$, the local displacements of the optimized topology are used to compute additional information about the problem, such as the VM stress field $\boldsymbol{x}_\mathtt{VM}$, and the dominant principal stress field $\boldsymbol{x}_\mathtt{TC}$, hereby denoted TC stress field. This latter refers to the tension or compression zones, made evident by the principal stress with greatest magnitude at each element of the domain. 

It is worth noting that both $\boldsymbol{x}_\mathtt{VM}$ and $\boldsymbol{x}_\mathtt{TC}$ are computed assuming the reference value of the elastic modulus $E_0$ everywhere. Clearly, these fields have no rigorous physical meaning since each element of the converged optimal solution is characterized by an elastic modulus weighted by the corresponding density value, according to \eq\eqref{eq:youngs_modulus_1}. The reason behind the choice of using $E_0$ for the entire design domain is due to the possibility of computing additional ``combined'' stress fields, as an element-wise product between either $\boldsymbol{x}_\mathtt{VM}$ or $\boldsymbol{x}_\mathtt{TC}$ and the optimal topology $\boldsymbol{x}$. The resulting combined stress fields not only recover physical meaning but, as shown later in \sez\ref{sec:experiments}, also allow for a better characterization of thin and intricate connection elements at prediction stage than that obtained for the density field alone.

\subsection{Real-time topology optimization model}
\label{sec:ML_formulation}

\begin{figure}[h!]
    \centering
    \includegraphics[width=13cm]{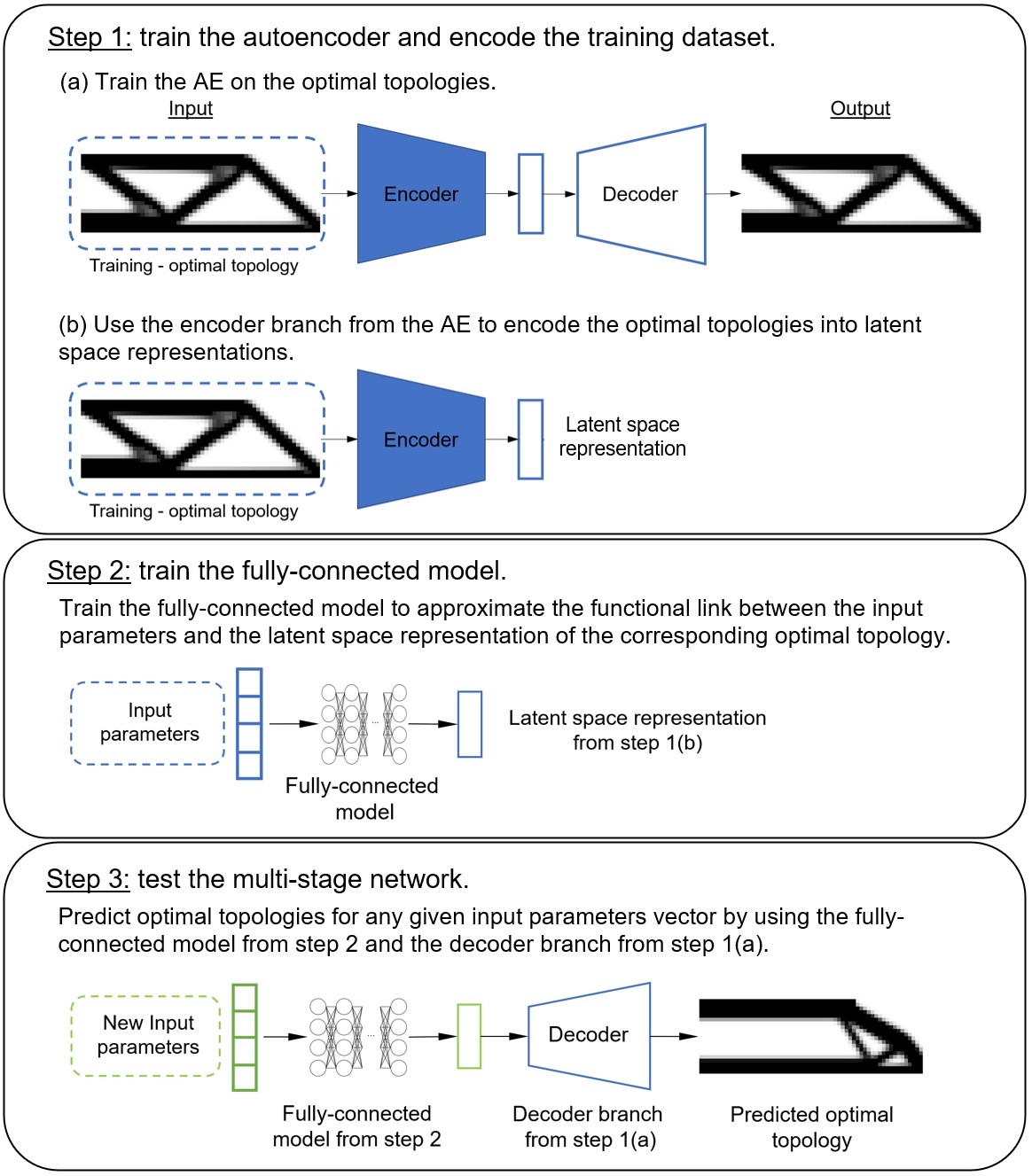}
    \caption{Scheme of the proposed deep learning pipeline.}
    \label{fig:ML_2}
\end{figure}

Starting from the FE-based optimal topologies, generated as detailed in the previous section and collected in the training dataset $\boldsymbol{D}$ according to \eq\eqref{eq:dataset}, this section describes the proposed DL-based strategy to approximate optimal designs in real-time for any input parameters vector $\boldsymbol{p}$. The proposed DL framework is schematized in \fig\ref{fig:ML_2} into three sequential steps. In the following, the three steps are detailed with reference to the prediction of optimal topologies, but the same also applies for the fields of VM stress $\boldsymbol{x}_\mathtt{VM}$ and dominant principal stress $\boldsymbol{x}_\mathtt{TC}$ without modifications. The three steps involve, respectively: (i) the training of an AE on optimal topologies, to encode them into their latent representation; (ii) the training of a fully-connected model to surrogate the functional link between the parameters underlying the loading and boundary conditions, and the latent representation of the corresponding optimal topology; (iii) the prediction of optimal topologies in real-time for any (unseen) parameters vector $\boldsymbol{p}$ by combining the fully-connected model and the AE decoder branch. Since the proposed strategy predicts an optimal topology for the prescribed design conditions without an iterative form finding process, it can be classified as part of the subgroup of machine learning-aided TO procedures termed \textit{non-iterative} methods.

\paragraph{Step 1: Train the AE.} A possible low-dimensional representation $\mathbf{h}\in\mathbb{R}^{N_L}$ of an optimal topology $\boldsymbol{x}$ in a low-dimensional space of size $N_L$, with $ N_L \ll N$, is learned by training an AE on the training dataset $\boldsymbol{D}$. The learnable components are the encoder $\mathcal{NN}_\mathtt{ENC}$ and decoder $\mathcal{NN}_\mathtt{DEC}$ branches of the AE, respectively providing the low-dimensional representation  $\mathbf{h}$, and the approximated reconstruction of the input topology $\overline{\boldsymbol{x}}\in\mathbb{R}^{N_e}$ obtained from $\mathbf{h}$, as follows:
\begin{align}
\mathbf{h}(\boldsymbol{x})=\mathcal{NN}_\mathtt{ENC}(\boldsymbol{x}(\boldsymbol{p})),\\
\overline{\boldsymbol{x}}(\mathbf{h})=\mathcal{NN}_\mathtt{DEC}(\mathbf{h}(\boldsymbol{x})).
\end{align}
The key component that links $\mathcal{NN}_\mathtt{ENC}$ and $\mathcal{NN}_\mathtt{DEC}$ is the so-called bottleneck layer representing the low-dimensional latent space of size $N_L$. This has a much smaller dimension than the input and output AE layers, thus forcing the data through a compressed representation while attempting to recreate the input as closely as possible at the output. 

During training (refer to ``step 1(a)'' in \fig\ref{fig:ML_2}), the weights $\mathbf{\Omega}_\mathtt{AE}$ parametrizing the AE are optimized in an unsupervised fashion minimizing one of the following loss functions, depending on whether the processed data pertain to density distributions or stress fields:
\begin{equation}
\mathcal{L}_\mathtt{AE}(\boldsymbol{\Omega}_\mathtt{AE},\boldsymbol{D}) =
\left\{
    \begin{array}{ll}
    -\frac{1}{I}\sum^{I}_{i=1} \boldsymbol{x}_i(\boldsymbol{p}_i)\text{log}(\overline{\boldsymbol{x}}_i(\mathbf{h}_i)) + (\boldsymbol{1} - \boldsymbol{x}_i)\text{log}(\boldsymbol{1}-\overline{\boldsymbol{x}}_i(\mathbf{h}_i)),& \textup{for $\boldsymbol{x}$},\smallskip\\
    \frac{1}{I}\sum^{I}_{i=1} \lVert\boldsymbol{x}_i(\boldsymbol{p}_i) - \mathcal{NN}_\mathtt{DEC}(\mathcal{NN}_\mathtt{ENC}(\boldsymbol{x}_i))\rVert_{2}^2,& \textup{for $\boldsymbol{x}_\mathtt{TC}$ and $\boldsymbol{x}_\mathtt{VM}$},
    \end{array}
    \right.
\label{eq:AE_loss}
\end{equation}
which are two loss functions typically exploited for training autoencoders \cite{book:DL_book}. 

The former is the probabilistic binary cross-entropy between the ground truth target label $\boldsymbol{x}(\boldsymbol{p})$ and its estimated counterpart $\overline{\boldsymbol{x}}(\mathbf{h})$. This serves to train the AE model in the case of density field data. The two classes refer to fully dense material and void elements, respectively, with the assignment of each element to one class or the other determined by a suitable discretization of the continuous $[0,1]$ range of values within which $\boldsymbol{x}$ can take on values. The binary cross-entropy loss function has also been employed, for instance, in the training of generative adversarial networks to predict density distributions in \cite{Behzadi2022} and in \cite{Nie2021}. The second entry in \eq\eqref{eq:AE_loss} is a standard mean squared error-like loss function. This loss function is utilized to train the AE model to reconstruct the input at the output, in both the cases of VM and TC stress fields data. The reason for considering a different loss function for the density field is that density data typically follow a binary distribution due to the use of the SIMP procedure for their generation, making them more suitable to be treated within a classification framework rather than a regression one. In both cases, minimizing $\mathcal{L}_\mathtt{AE}$ is equivalent to maximizing the likelihood of $\mathbf{D}$ on the parameters $\boldsymbol{\Omega}_\mathtt{AE}$.

Once the AE is trained, the optimal topologies in $\boldsymbol{D}$ are mapped onto the low-dimensional feature space (refer to ``step 1(b)'' in \fig\ref{fig:ML_2}), once and for all, to provide 
\begin{equation}
\boldsymbol{D}_L = \{(\mathbf{h}(\boldsymbol{x}), \mathbf{h}(\boldsymbol{x}_\mathtt{VM}), \mathbf{h}(\boldsymbol{x}_\mathtt{TC}), \boldsymbol{p})_i \}^{I}_{i=1},
\label{eq:dataset_low}
\end{equation}
which collects the latent representations $\mathbf{h}$ of the training data and the relative labels, in terms of the parameters vector $\boldsymbol{p}$ ruling the boundary and loading conditions.

\paragraph{Step 2: Train the fully-connected model.} The second step of the proposed deep learning pipeline consists of training a fully-connected model that approximates the functional link between the parametric input space underlying $\boldsymbol{p}$ and the low-dimensional feature space described by $\mathcal{NN}_\mathtt{ENC}$, as follows:
\begin{equation}
\overline{\mathbf{h}}=\mathcal{NN}_\mathtt{FC}(\boldsymbol{p}),
\label{eq:FC}
\end{equation}
where $\overline{\mathbf{h}}\in\mathbb{R}^{N_L}$ denotes the approximation to the low-dimensional features obtained from $\mathcal{NN}_\mathtt{ENC}$ for the corresponding optimal topology.

During training, the weights $\mathbf{\Omega}_\mathtt{FC}$ parametrizing $\mathcal{NN}_\mathtt{FC}$ are tuned by minimizing the following loss function:
\begin{equation}
\mathcal{L}_\mathtt{FC}(\boldsymbol{\Omega}_\mathtt{FC},\mathbf{D}_L) =\frac{1}{I}\sum^{I}_{i=1}\lVert \mathbf{h}_i(\boldsymbol{x}_i) - \mathcal{NN}_\mathtt{FC}(\boldsymbol{p}_i)\rVert_{2}^2,
\label{eq:FC_loss}
\end{equation}
which provides a measure of the distance between the target low dimensional features vector $\mathbf{h}(\boldsymbol{x})$, obtained through the feature extractor $\mathcal{NN}_\mathtt{ENC}$, and its approximated counterpart $\overline{\mathbf{h}}=\mathcal{NN}_\mathtt{FC}(\boldsymbol{p})$, provided by the fully-connected model.

\paragraph{Step 3: Predict optimal topologies in real-time.} Once both the AE and fully-connected models are trained, the optimal topology (or, equivalently, the fields of VM stress and dominant principal stress) for any given input parameters vector $\boldsymbol{p}^*\in\mathbb{R}^\mathtt{N_{par}}$, unseen during training, is predicted in real-time through the combined use of the fully-connected model and the decoder branch of the AE, as follows: 
\begin{equation}
\boldsymbol{x}^* = \mathcal{NN}_\mathtt{DEC}(\mathcal{NN}_\mathtt{FC}(\boldsymbol{p}^*)),
\label{eq:Prediction}
\end{equation}
where $\boldsymbol{x}^*\in\mathbb{R}^{N_e}$ denotes the neural network approximation to the optimal design for a given vector of input parameters $\boldsymbol{p}^*$ unseen during training.

\section{Numerical experiments}
\label{sec:experiments}

In this section, we assess the proposed strategy against two test cases of increasing structural complexity, respectively related to the topology optimization of a 2D MBB beam and of a 3D bridge. For the two case studies, the dataset generation is carried out in the \texttt{MATLAB} environment, respectively using the implementation codes \cite{Sigmund2001} and \cite{Ferrari2020}. All the computations are carried out on a PC featuring an \texttt{Intel\textsuperscript{\textregistered} Core\textsuperscript{TM} i7-8650U} CPU @ 1.90 GHz and 16 GB RAM. The DL models are implemented through the \texttt{TensorFlow}-based \texttt{Keras} API, and trained on a \texttt{Google Colab} virtual machine featuring two CPUs with x86-64 processors and 11 GB RAM. The implementation details of the DL models are reported in Appendix~\ref{sec:implementation}.

\subsection{2D MBB beam}
\label{sec:2D_MBB}

\begin{figure}[h!]
    \centering
    \includegraphics[width=10cm]{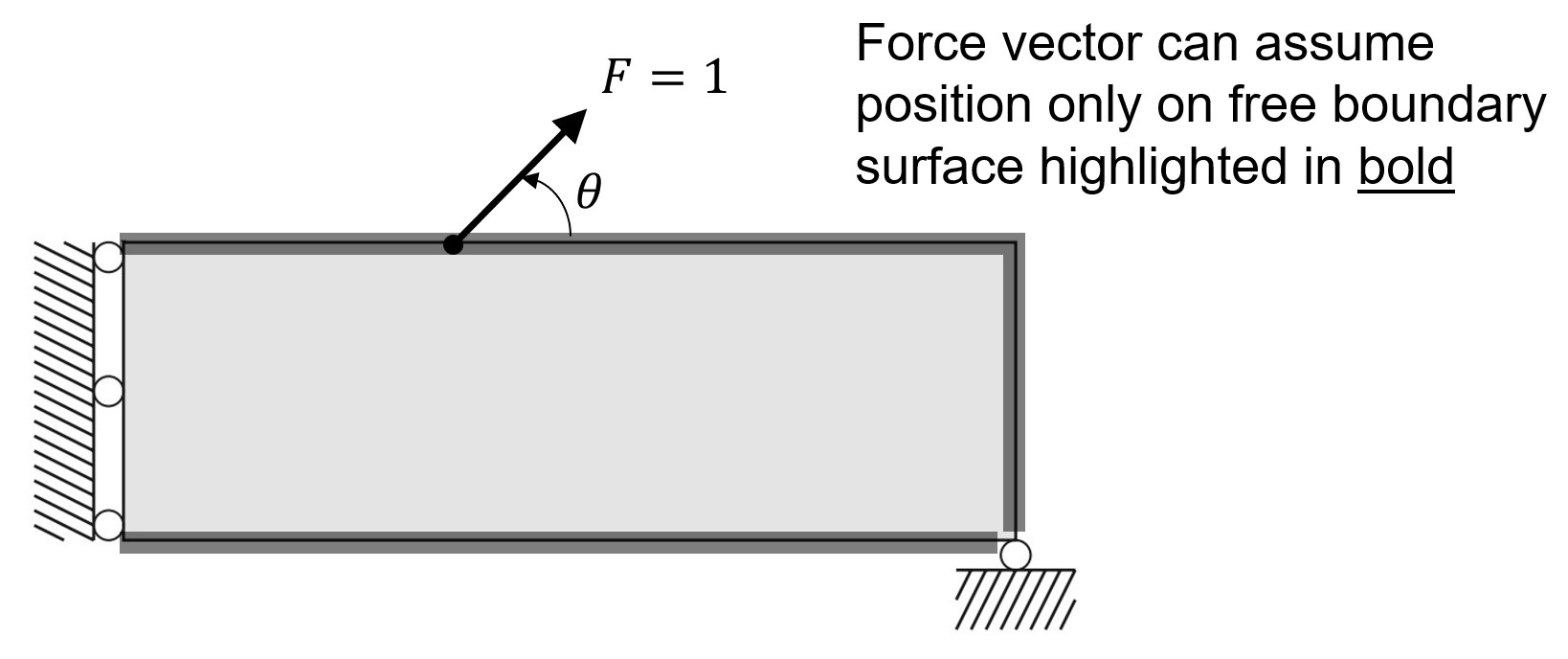}
    \caption{MBB beam. Half-beam with details of an exemplary loading condition and allowed locations of the force vector application point.}
    \label{fig:2D_MBB_1}
\end{figure}

The first test case involves the 2D MBB beam depicted in \fig\ref{fig:2D_MBB_1}. By exploiting symmetry, half of the beam is considered to reduce the computational burden. The half-beam has the bottom-right corner restrained in the vertical direction, while the left edge is restrained in the horizontal direction and can slide vertically. The rectangular domain is discretized using linear quadrilateral elements, with a FE mesh obtained from a $120\times40$ regular grid, resulting in $N_e=4800$ finite elements. The structure is loaded by a unit concentrated load, whose position and direction are parametrized by means of $\mathtt{N_{par}}=3$ parameters, collected in the vector $\boldsymbol{p}=\{x_{F}, y_{F}, \theta\}$. The three parameters describe the force position along the $x$-axis and $y$-axis as $x_{F} \in [0,60]$ and $y_{F} \in [0,20]$, and the orientation angle of the force in degrees as $\theta\in [0,90]$, respectively. In particular, the force vector can be applied in any node of the free boundary surface, as shown in \fig\ref{fig:2D_MBB_1}.

Concerning the topology optimization setup, the target volume fraction is prescribed as $f=0.5$, the penalization factor is set as $p=3$, and the filter radius is equal to $r_{min} = 1.5$. At the end of each optimization loop, the optimal topology $\boldsymbol{x}$, as well as the related stress fields $\boldsymbol{x}_\mathtt{VM}$ and $\boldsymbol{x}_\mathtt{TC}$, are provided to populate $\boldsymbol{D}$, according to \eq\eqref{eq:dataset}. For the case at hand, the DL models $\mathcal{NN}_\mathtt{ENC}$, $\mathcal{NN}_\mathtt{DEC}$ and $\mathcal{NN}_\mathtt{FC}$ are trained exploiting $I=2500$ data instances, generated for different values of the input parameters $\boldsymbol{p}$ sampled via latin hypercube rule.

\paragraph{Results.}
For the present case study, the testing data consist of $500$ additional data instances not seen during training, and exploited to assess the performance of the proposed strategy. The first quantitative evaluation of the benefit of the proposed framework is given by the comparison between the computational time taken to generate an optimal topology using the SIMP-based algorithm and through the combined use of $\mathcal{NN}_\mathtt{FC}$ and $\mathcal{NN}_\mathtt{DEC}$. As shown in \tab\ref{tab:2DMBB_time}, the traditional SIMP-based procedure takes on average $12.00$ seconds to provide an optimized topology. The proposed method instead requires the preliminary phases of dataset creation and training, which take $8.10$ hours and $0.44$ hours, respectively. However, once trained, the DL model provides optimized topologies almost instantaneously. Concerning the prediction of the VM or the TC stress field, this does not entail an additional computational burden for the dataset creation. Nevertheless, there is an additional model to train, yielding a doubled training time of $0.88$ hours. The same applies to the prediction run-time, which doubles to $0.16$ seconds due to the necessity of evaluating two models.

\begin{table}[t!]
\caption{MBB beam. Average computational run-time to generate optimal topologies using the SIMP-based procedure and the proposed framework.}
\centering
\small
\begin{tabular}{cccc}
\toprule
\mbox{-} & \mbox{SIMP Method} & \multicolumn{1}{p{3.5cm}}{\centering \mbox{Proposed method:} \\ \mbox{$\boldsymbol{x}$ only}} & \multicolumn{1}{p{3.5cm}}{\centering \mbox{Proposed method:} \\ \mbox{$\{\boldsymbol{x},\boldsymbol{x}_\mathtt{VM}\}$ or $\{\boldsymbol{x},\boldsymbol{x}_\mathtt{TC}\}$}}\\
\toprule
\mbox{Dataset creation} & \mbox{-} & \mbox{$8.10$ hours} & \mbox{$8.10$ hours} \\
\mbox{Training time} & \mbox{-} & \mbox{$0.44$ hours} & \mbox{$0.88$ hours} \\
\multicolumn{1}{p{3cm}}{\centering \mbox{Average run-time}} & \mbox{$12.00$ seconds} & \mbox{$0.08$ seconds} & \mbox{$0.16$ seconds} \\
\bottomrule
\end{tabular}
\label{tab:2DMBB_time}
\end{table}

\begin{figure}[t!]
    \centering
    \includegraphics[width=11.5cm]{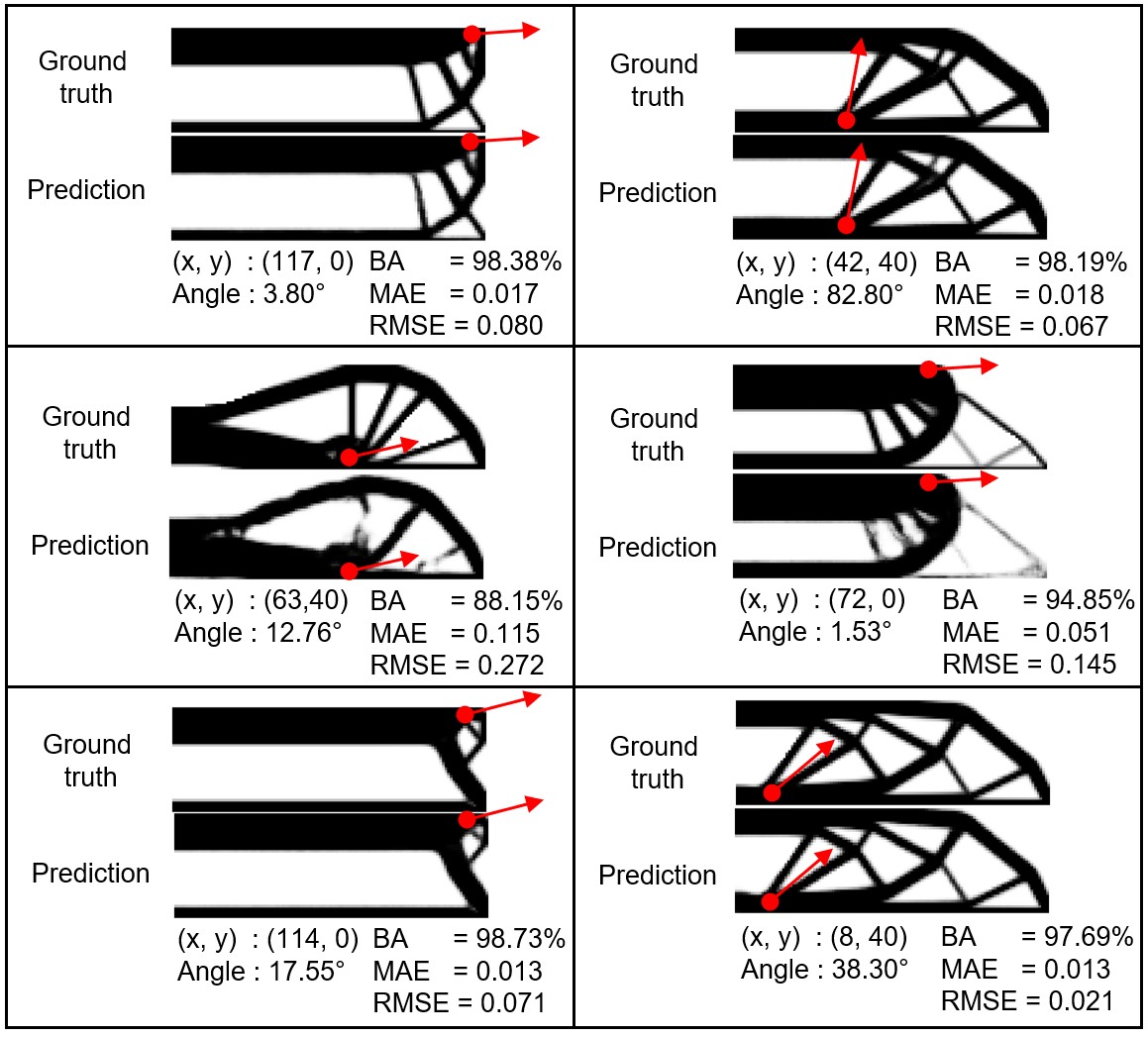}
    \caption{MBB beam. Exemplary comparisons of optimal topologies obtained using the SIMP-based procedure and through the proposed framework. Results in each box are obtained for varying locations and orientations of the acting force. For each case, the accuracy of the predicted topology is measured in terms of binary accuracy (BA), mean absolute error (MAE) and root mean squared error (RMSE).}
    \label{fig:2DMBB_results_topopt}
\end{figure}

To demonstrate the predictive capabilities achieved by the trained multi-stage DL model, some exemplary optimal topologies provided through the combined use of $\mathcal{NN}_\mathtt{FC}$ and $\mathcal{NN}_\mathtt{DEC}$ are reported in \fig\ref{fig:2DMBB_results_topopt}. Each box reports the target ground truth topology, optimized with the SIMP-based procedure, together with the corresponding approximation, predicted using the proposed model. These topologies are attained for varying input parameters, chosen to showcase the obtained results for a variety of design forms. The accuracy of the predicted density field $\boldsymbol{x}$ is assessed in terms of binary accuracy (BA), mean absolute error (MAE) and root mean squared error (RMSE), which are reported below each test case. The BA indicator is useful to understand the predictive capabilities of a model in the presence of binary data. It  measures the ratio between the number of correct classifications to the total number of classifications, with reference to a binary classification problem. In the case of density field data, the two classes refer to fully dense material and void elements, respectively, with the value of each element assigned to one class or the other according to a suitable discretization of the continuous values range of interest. The MAE is a common metric used to assess the accuracy of a regression model. It measures the average magnitude of the differences between the predicted and actual values, providing a straightforward interpretation of the prediction accuracy that is not very sensitive to outliers. Similarly, the RMSE measures the average magnitude of the differences between predicted and actual values, taking into account both the magnitude and the direction of errors. It is particularly sensitive to outliers due to the squaring operation, making it an useful metric to use in conjunction with the MAE.

\begin{figure}[t!]
\centering
\captionsetup[subfigure]{justification=centering}\subfloat[\label{fig:2D_VM}]{\includegraphics[width=.8\textwidth]{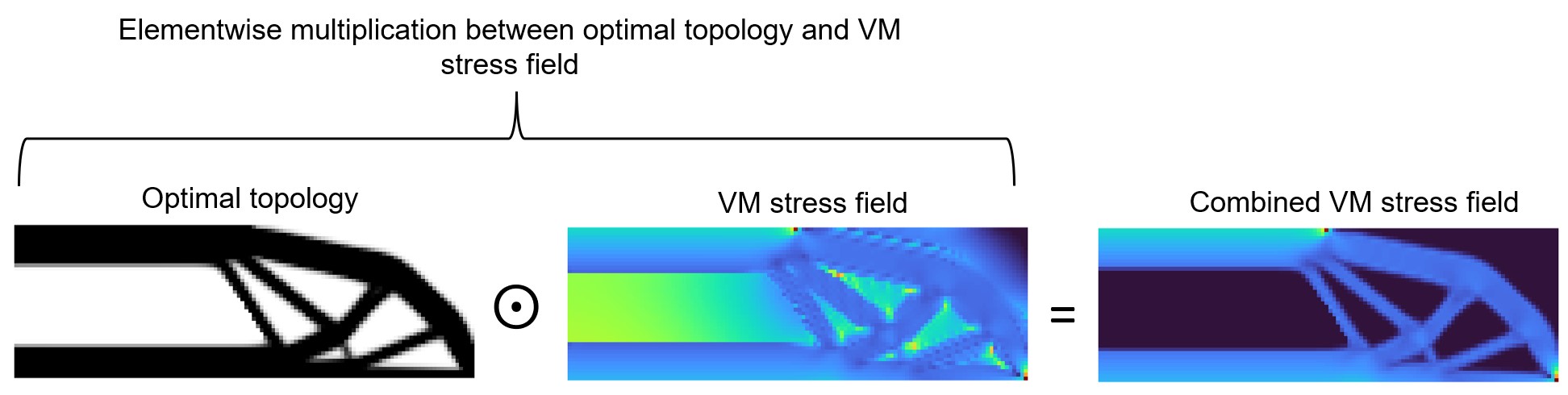}}\\
\subfloat[\label{fig:2D_TC}]{\includegraphics[width=.8\textwidth]{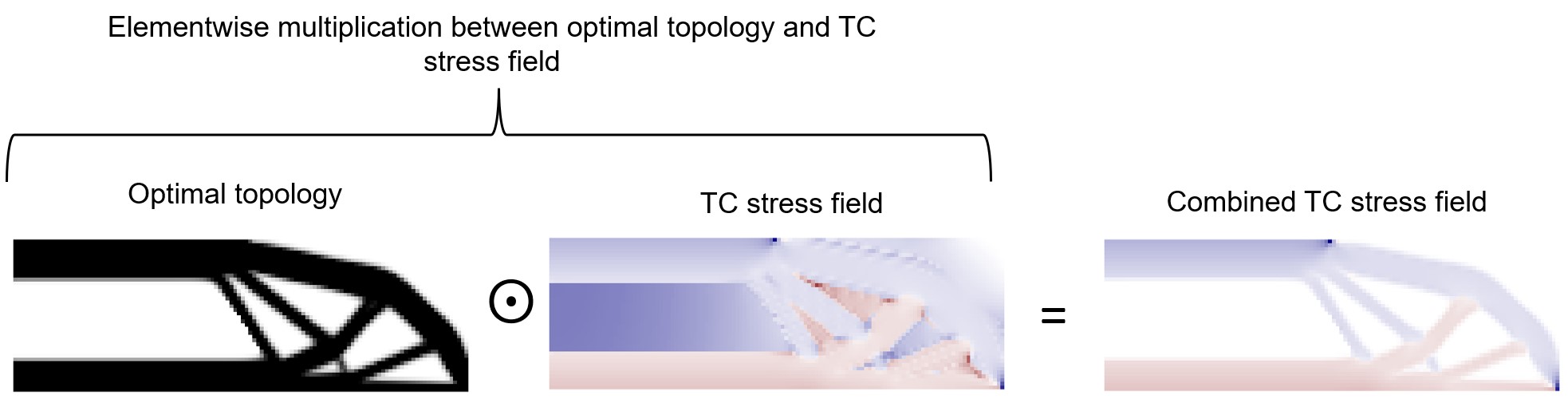}}
\caption{MBB beam. Scheme of the combined stress fields computation as an element-wise multiplication between (a) the optimal topology and the corresponding VM stress field, (b) the optimal topology and the corresponding TC stress field.\label{fig:2D_VMandTC}}
\end{figure}

Before presenting the results for the prediction of the VM and TC stress fields, we briefly comment on the procedure to compute the relative combined fields. As described in \sez\ref{sec:TO_formulation}, the prediction of $\boldsymbol{x}_\mathtt{VM}$ and  $\boldsymbol{x}_\mathtt{TC}$ provides an approximation to the VM and TC stress fields, respectively, under the assumption that the entire design domain is characterized by the reference value of the elastic modulus $E_0$. To recover physical consistency and improve the characterization of thin and intricate connection elements in the prediction of $\boldsymbol{x}$ (see \fig\ref{fig:2DMBB_results_topopt}), the predictions of the two stress fields are projected onto the optimal topology, as shown in \fig\ref{fig:2D_VMandTC}. This is obtained as an element-wise multiplication between either $\boldsymbol{x}^*$ and $\boldsymbol{x}_\mathtt{VM}^*$ or  $\boldsymbol{x}^*$ and $\boldsymbol{x}_\mathtt{TC}^*$. In the resulting combined stress fields, elements characterized by a zero density of material do not contribute when multiplied with the corresponding VM or TC stress value, while elements with a density of material equal or almost equal to one retain a VM or TC stress value as previously computed.

\begin{figure}[h!]
    \centering
    \includegraphics[width=15cm]{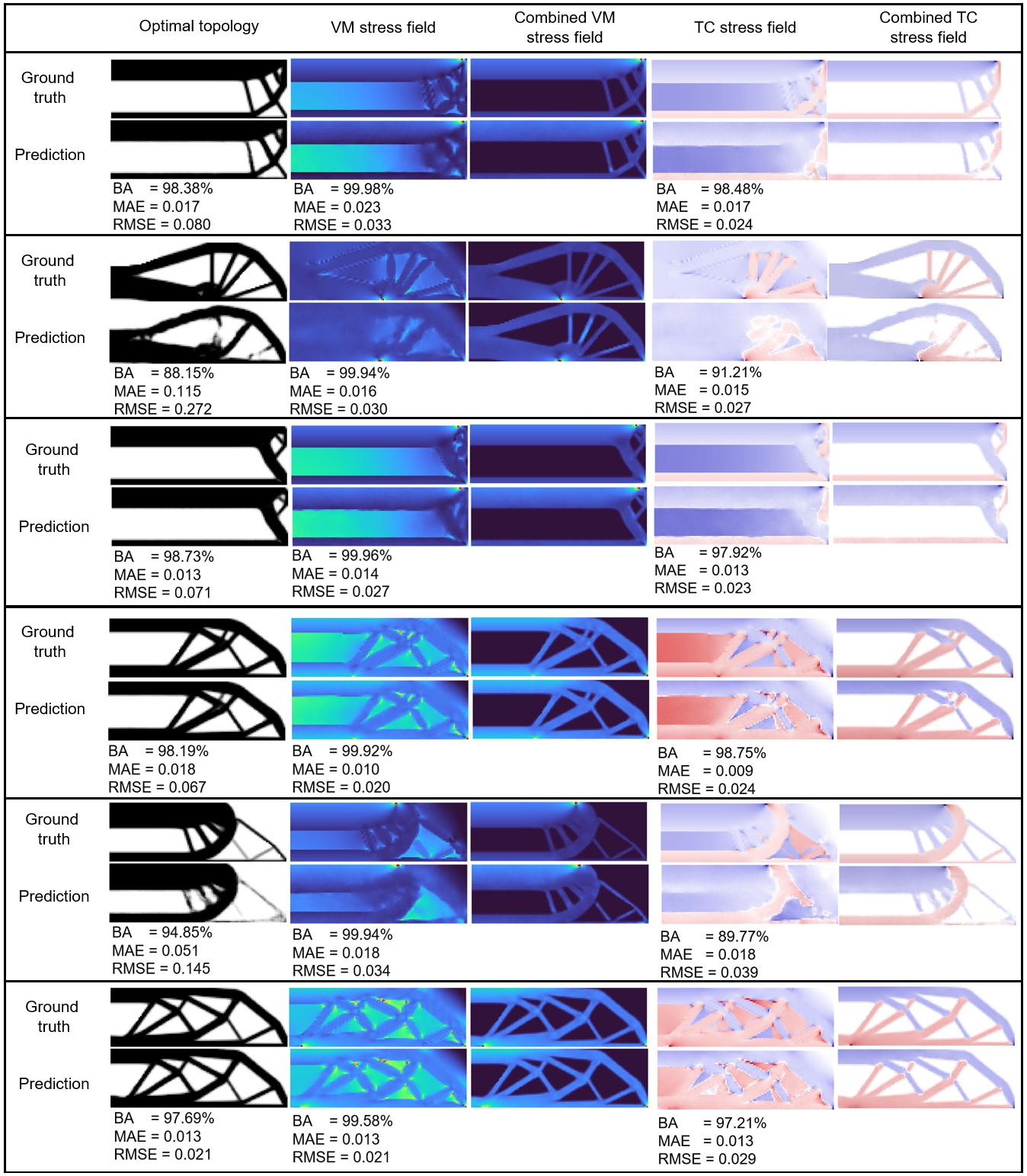}
    \caption{MBB beam. Exemplary comparisons of optimal topologies, VM and TC stress fields, and the corresponding combined fields, obtained using the SIMP-based procedure and through the proposed framework. Results in each box are obtained for varying locations and orientations of the acting force. For each case, the accuracy of the predictions is measured in terms of binary accuracy, mean absolute error and root mean squared error.}
    \label{fig:2DMBB_results_VM_TC}
\end{figure}

Additional results regarding the prediction of the $\boldsymbol{x}_\mathtt{VM}$ and $\boldsymbol{x}_\mathtt{TC}$ stress fields, as well as the relative combined fields, are shown in \fig\ref{fig:2DMBB_results_VM_TC}. For each test case, the top row reports the ground truth results, obtained with the SIMP-based procedure, while the bottom row reports the corresponding approximations, obtained through the proposed DL model. The first column presents the same optimal topologies as shown in \fig\ref{fig:2DMBB_results_topopt}, while the next four columns show the $\boldsymbol{x}_\mathtt{VM}$ and $\boldsymbol{x}_\mathtt{TC}$ stress fields, together with their combined counterpart. For each test case, the BA, MAE, and RMSE accuracy metrics are reported below the three predicted fields. An overall quantitative assessment of the predictive capabilities of the proposed framework is finally shown in \tab\ref{tab:2DMBB_accuracy}, reporting the three performance indicators for the predicted $\boldsymbol{x}$, $\boldsymbol{x}_\mathtt{VM}$ and $\boldsymbol{x}_\mathtt{TC}$ fields, averaged over the $500$ testing instances.

A comparative study to investigate the impact of specific choices in the architecture of DL models on the accuracy achieved in approximating optimal topologies for the present case study is reported in Appendix~\ref{sec:comparative_exp}. These analyses are performed by training new autoencoders and the fully-connected models with varying numbers of layers, and then testing each possible combination for the combined use of the fully-connected model and the decoding branch of the autoencoder. The outcomes of these additional experiments are eventually compared with the results obtained using the reference architectures described in Appendix~\ref{sec:implementation}.

\begin{table}[b!]
\caption{MBB beam. Accuracy of the proposed DL model against the ground truth SIMP-based procedure. Results are reported in terms of binary accuracy, mean absolute error and root mean squared error, all averaged over $500$ testing instances.}
    \centering
    \begin{tabular}{cccc}
    \toprule
    \mbox{Accuracy metric} & \mbox{$\boldsymbol{x}$ density field} &  \mbox{$\boldsymbol{x}_\mathtt{VM}$ stress field} & \mbox{$\boldsymbol{x}_\mathtt{TC}$ stress field}\\
    \toprule
    \mbox{BA}               & \mbox{$96.46\%$}    & \mbox{$99.29\%$}        & \mbox{$95.42\%$}\\
    \mbox{MAE}              & \mbox{$0.035$}      & \mbox{$0.018$}          & \mbox{$0.013$}\\
    \mbox{RMSE}              & \mbox{$0.107$}      & \mbox{$0.030$}          & \mbox{$0.024$}\\
    \bottomrule
    \end{tabular}
    \label{tab:2DMBB_accuracy}
\end{table}

\paragraph{Discussion of the results.}
The results obtained for the present case study are in general
fairly good. The topology optimization predictions yield accuracy metrics values averaged over the testing data equal to $96.46\%$ for the BA, $0.035$ for the MAE, and $0.107$ for the RMSE, respectively. The $\boldsymbol{x}_\mathtt{VM}$ predictions score $99.29\%$, $0.018$, and $0.030$, while the $\boldsymbol{x}_\mathtt{TC}$ predictions score $95.42\%$, $0.013$, and $0.024$, respectively, for the same accuracy metrics. To comment on the obtained results, some key aspects can be highlighted as follows.

\begin{itemize}
  \item The $\boldsymbol{x}$ predictions performed the worst in terms of the MAE and RMSE accuracy metrics when compared to the $\boldsymbol{x}_\mathtt{VM}$ or to the $\boldsymbol{x}_\mathtt{TC}$ stress field predictions. This is due to the almost $0-1$ binary distribution of the density field induced by the SIMP-based procedure. This aspect implies that an erroneous prediction of a void or a dense element incurs a high penalty on the MAE and RMSE scores.
  \item Image patterns related to thin and intricate connection elements are more difficult to learn and predict, and often result in a degradation of all the performance indicators. For instance, in the second and fifth test cases shown in \fig\ref{fig:2DMBB_results_VM_TC}, this aspect moderately spoils the obtained results compared to the other test cases. This behavior stems from the fact that such connection elements are typically subtle and do not occur regularly in the training dataset, thus reducing the confidence DL models have in dealing with them. However, this aspect is mitigated in the prediction of the $\boldsymbol{x}_\mathtt{VM}$ and $\boldsymbol{x}_\mathtt{TC}$ stress fields, as testified by the reduced attained values of MAE and RMSE. This is once again due to the distribution of values characterizing the two stress fields, which significantly differs from the $0-1$ binary distribution characterizing $\boldsymbol{x}$, and for which incorrect predictions do not deviate markedly from the ground truth.
  \item In most of the cases, the combined stress fields address the incorrect or confused prediction of thin connection elements displayed by the density field. As shown in the second and fifth test cases of \fig\ref{fig:2DMBB_results_VM_TC}, the element-wise product between either $\boldsymbol{x}^*$ and $\boldsymbol{x}_\mathtt{VM}^*$ or $\boldsymbol{x}^*$ and $\boldsymbol{x}_\mathtt{TC}^*$ allows for a better characterization of thin and intricate connection elements than that provided by $\boldsymbol{x}^*$. This is because these regions are characterized by non-zero, although relatively small, predicted density values, as well as by predicted stress values higher than the ground truth. Weighting the predicted stress fields by the corresponding values of the predicted density field thus allows a better characterization of these connection elements.
  \item As shown in \fig\ref{fig:2DMBB_results_VM_TC}, stress concentration regions are clearly visible near the load application point. Other areas of significant stress concentration can be observed close to the restraints; however when compared to the stress concentration region induced by the load application, the color scaling reduces their visibility. While introducing passive domains could potentially induce artificial stress concentration regions, passive domains are not enforced in the present case study. Therefore, no additional areas of significant stress concentration are observed.
\end{itemize}

\subsection{3D bridge}
\label{sec:3D_bridge}

\begin{figure}[h!]
    \centering
    \includegraphics[width=11cm]{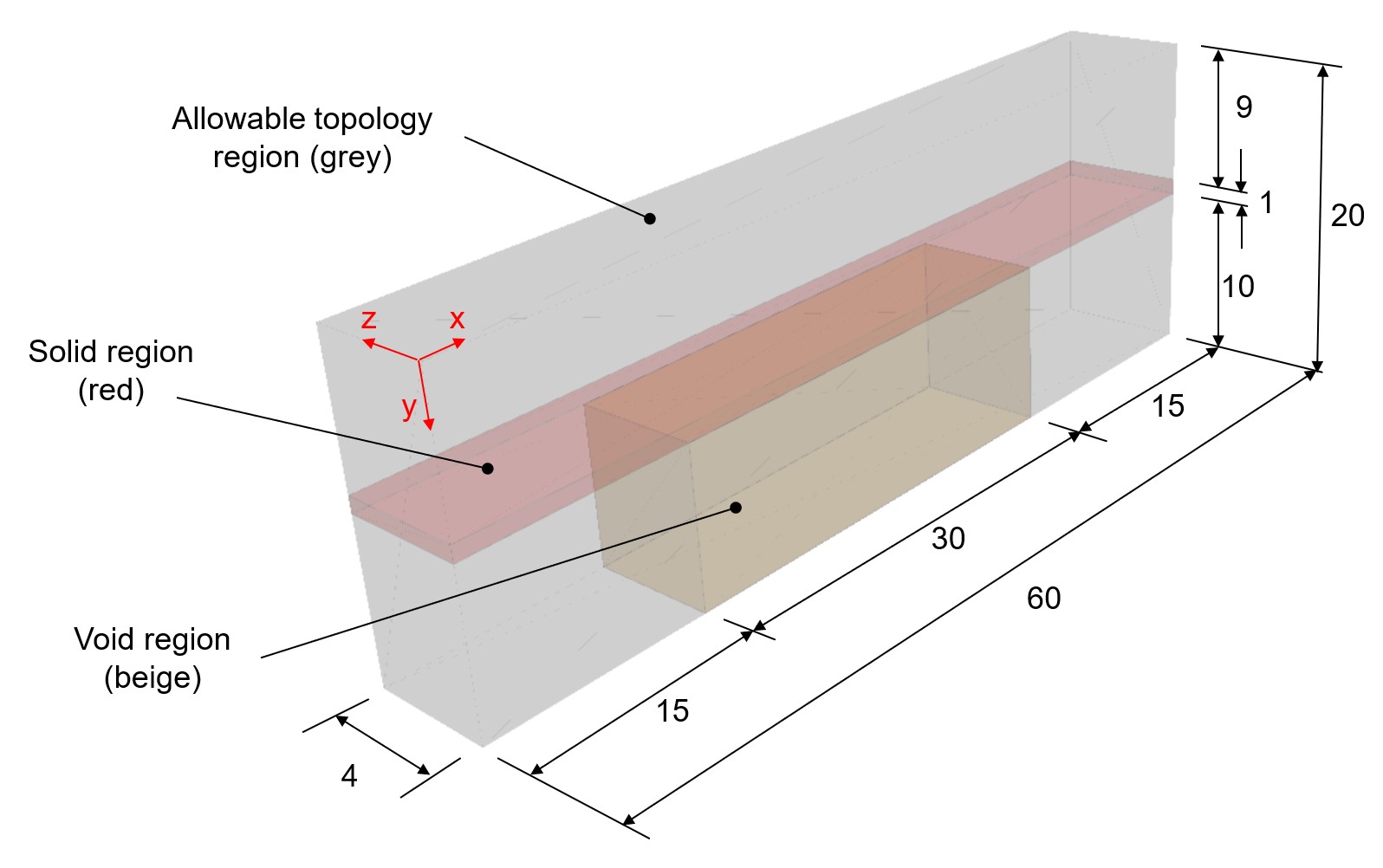}
    \caption{3D bridge. Half-bridge with details of the domain dimensions, and of the solid, void, and allowed topology regions.}
    \label{fig:3DBridge_regions}
\end{figure}

\begin{figure}[h!]
    \centering
    \includegraphics[width=9cm]{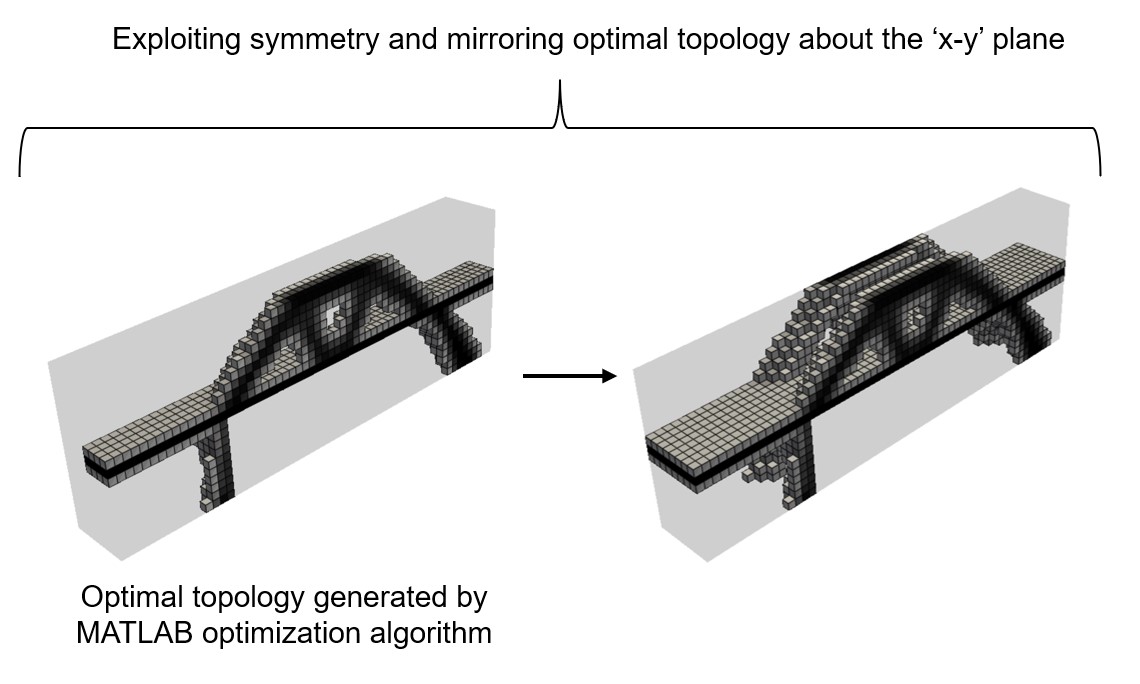}
    \caption{3D bridge. Mirroring the half-bridge domain to recover the full bridge configuration.}
    \label{fig:3DBridge}
\end{figure}

This second case study aims to assess the performance of the proposed TO framework in the more involved situation of the 3D bridge-like structure depicted in \fig\ref{fig:3DBridge_regions}. Also in this case, only one half of the target design domain is considered by exploiting the symmetry with respect to the $x-y$ plane at $z=0$. The full bridge configuration can be easily recovered as shown in \fig\ref{fig:3DBridge}. The domain is discretized using 8-node hexahedron elements, with a FE mesh obtained from a $60\times20\times4$ regular grid, resulting in $N_e = 4800$ finite elements. As highlighted in \fig\ref{fig:3DBridge_regions}, the deck zone (highlighted in red) is constrained to feature a full material density and has a unit thickness. The void region (highlighted in beige) located below the deck is modeled using passive elements and features a length of $30$ units. The topology is free to develop in the rest of the design domain (highlighted in grey). 

\begin{figure}[h!]
    \centering
    \includegraphics[width=11cm]{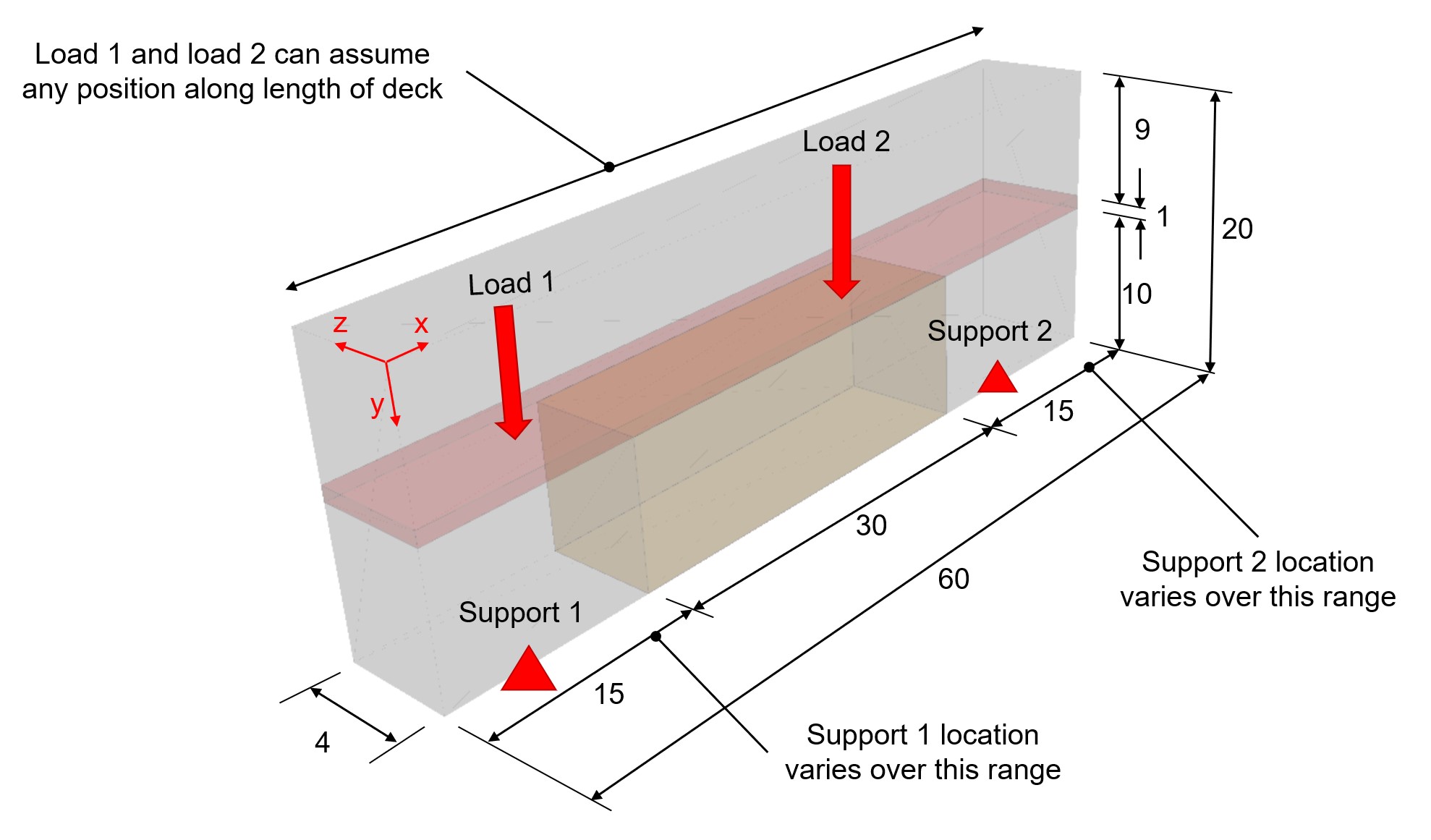}
    \caption{3D bridge case. Schematic representation of the parameters ruling the parametric input space underlying the SIMP-based procedure.}
    \label{fig:3DBridge_loading}
\end{figure}

The structure is loaded by two concentrated loads of unit magnitude applied on the deck, pointing vertically downward, and whose position along the $x$-axis is independently parametrized to vary along the length of the deck. Two other parameters are used to control the position of the supports along the $x$-axis in the two regions neighboring the central void, as shown in \fig\ref{fig:3DBridge_loading}. In total, the problem is parametrized by means of $\mathtt{N_{par}}=4$ parameters, collected as $\boldsymbol{p}=\{x_{F1}, x_{F2}, x_{S1}, x_{S2}\}$, and respectively describing the position of the two acting forces along the $x$-axis as $\{x_{F1}, x_{F2}\} \in [0,60]$, and the position of the two supports along the $x$-axis as $x_{S1} \in [0,15]$ and $x_{S2} \in [45,60]$.

\begin{figure}[h!]
\centering
\captionsetup[subfigure]{justification=centering}\subfloat[\label{fig:3DBridge_VM}]{\includegraphics[width=.8\textwidth]{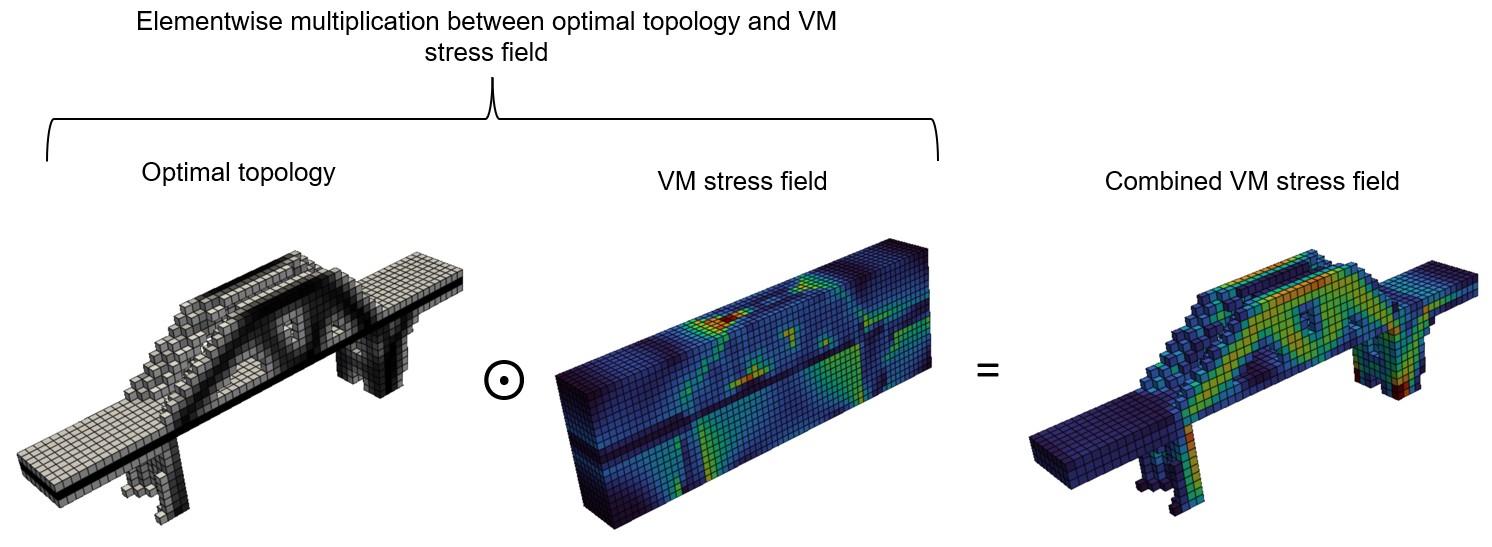}}\\
\subfloat[\label{fig:3DBridge_TC}]{\includegraphics[width=.8\textwidth]{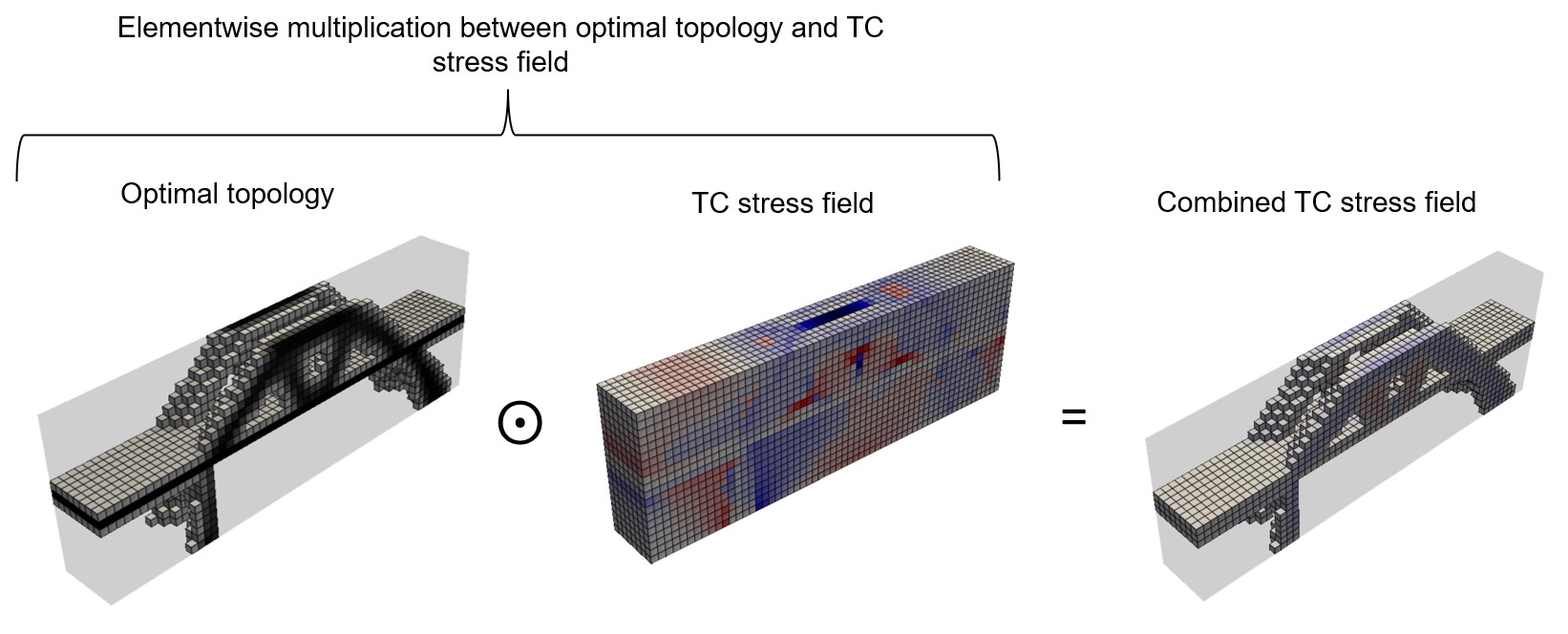}}
\caption{3D bridge. Scheme of the combined stress fields computation as an element-wise multiplication between (a) the optimal topology and the corresponding VM stress field, (b) the optimal topology and the corresponding TC stress field.\label{fig:3DBridge_VMandTC}}
\end{figure}

Concerning the topology optimization setup, the target volume fraction is prescribed as \mbox{$f = 0.12$}, the penalization factor is set as $p = 3$, and the filter radius is equal to $r_{min} = \sqrt{3}$. In this case, $I=2500$ training data instances are generated for varying values of the input parameters $\boldsymbol{p}$ sampled via latin hypercube rule, to populate $\boldsymbol{D}$ according to \eq\eqref{eq:dataset}. The masking procedure to compute the combined stress fields follows the same concept employed for the 2D MBB beam, as schematically illustrated in \fig\ref{fig:3DBridge_VMandTC}.

\paragraph{Results.}
For the present case, the testing data consist of 500 data instances generated through the SIMP-based procedure and not seen during training. The comparison between the computational time taken to generate optimal topologies using the SIMP-based procedure and the combined use of $\mathcal{NN}_\mathtt{FC}$ and $\mathcal{NN}_\mathtt{DEC}$ is reported in \tab\ref{tab:3DBridge_time}. The computational gain achieved by the proposed framework in predicting new topologies is even more remarkable than in the previous case study, yielding an average speed-up of about $287$ times. 

\begin{table}[h!]
\caption{3D bridge. Average computational run-time to generate optimal topologies using the SIMP-based procedure and the proposed framework.}
    \centering
    \begin{tabular}{cccc}
    \toprule
    \mbox{-} & \mbox{SIMP Method} & \multicolumn{1}{p{3.5cm}}{\centering \mbox{Proposed method:} \\ \mbox{$\boldsymbol{x}$ only}} & \multicolumn{1}{p{3.5cm}}{\centering \mbox{Proposed method:} \\ \mbox{$\{\boldsymbol{x},\boldsymbol{x}_\mathtt{VM}\}$ or $\{\boldsymbol{x},\boldsymbol{x}_\mathtt{TC}\}$}}\\
    \toprule
    \mbox{Dataset creation}                & \mbox{-}      & \mbox{$25.91$ hours}               & \mbox{$25.91$ hours}\\
    \mbox{Training time}                & \mbox{-}    & \mbox{$0.61$ hours}               & \mbox{$0.61$ hours}\\
    \multicolumn{1}{p{3cm}}{\centering \mbox{Average run-time}} & \mbox{$37.31$ seconds} & \mbox{$0.13$ seconds} & \mbox{$0.26$ seconds} \\
    \bottomrule
    \end{tabular}
    \label{tab:3DBridge_time}
\end{table}

\begin{figure}
    \centering
    \includegraphics[width=11.5cm]{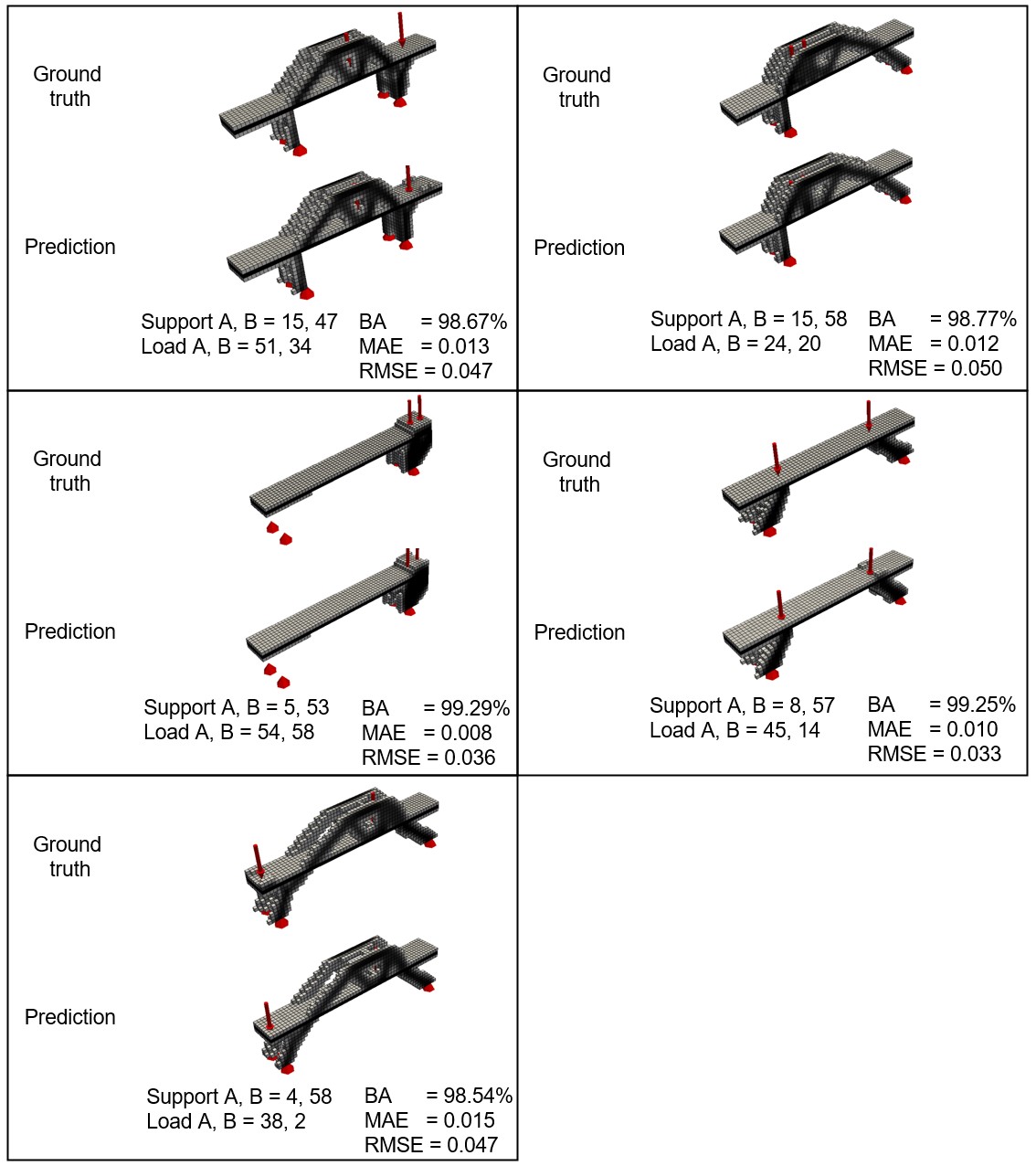}
    \caption{3D bridge. Exemplary comparisons of optimal topologies obtained using the SIMP-based procedure and through the proposed framework. Results in each box are obtained for varying positions of the acting forces and of the supports. For each case, the accuracy of the predicted topology is measured in terms of binary accuracy (BA), mean absolute error (MAE) and root mean squared error (RMSE).}
    \label{fig:3DBridge_results_topopt}
\end{figure}

Some optimal topologies obtained through the combined use of $\mathcal{NN}_\mathtt{FC}$ and $\mathcal{NN}_\mathtt{DEC}$ are shown in \fig\ref{fig:3DBridge_results_topopt}. Each box reports the target ground truth topology, obtained through the SIMP-based procedure, and the corresponding approximation, predicted with the proposed model. These topologies are provided for different combinations of the locations of the two acting forces along the deck and of the supports, which are overlaid on each topology and specified below each test case. The accuracy of the predicted density field is assessed in terms of the BA, MAE, and RMSE performance indicators, which are reported below each test case.

\begin{figure}
    \centering
    \includegraphics[width=.97\textwidth]{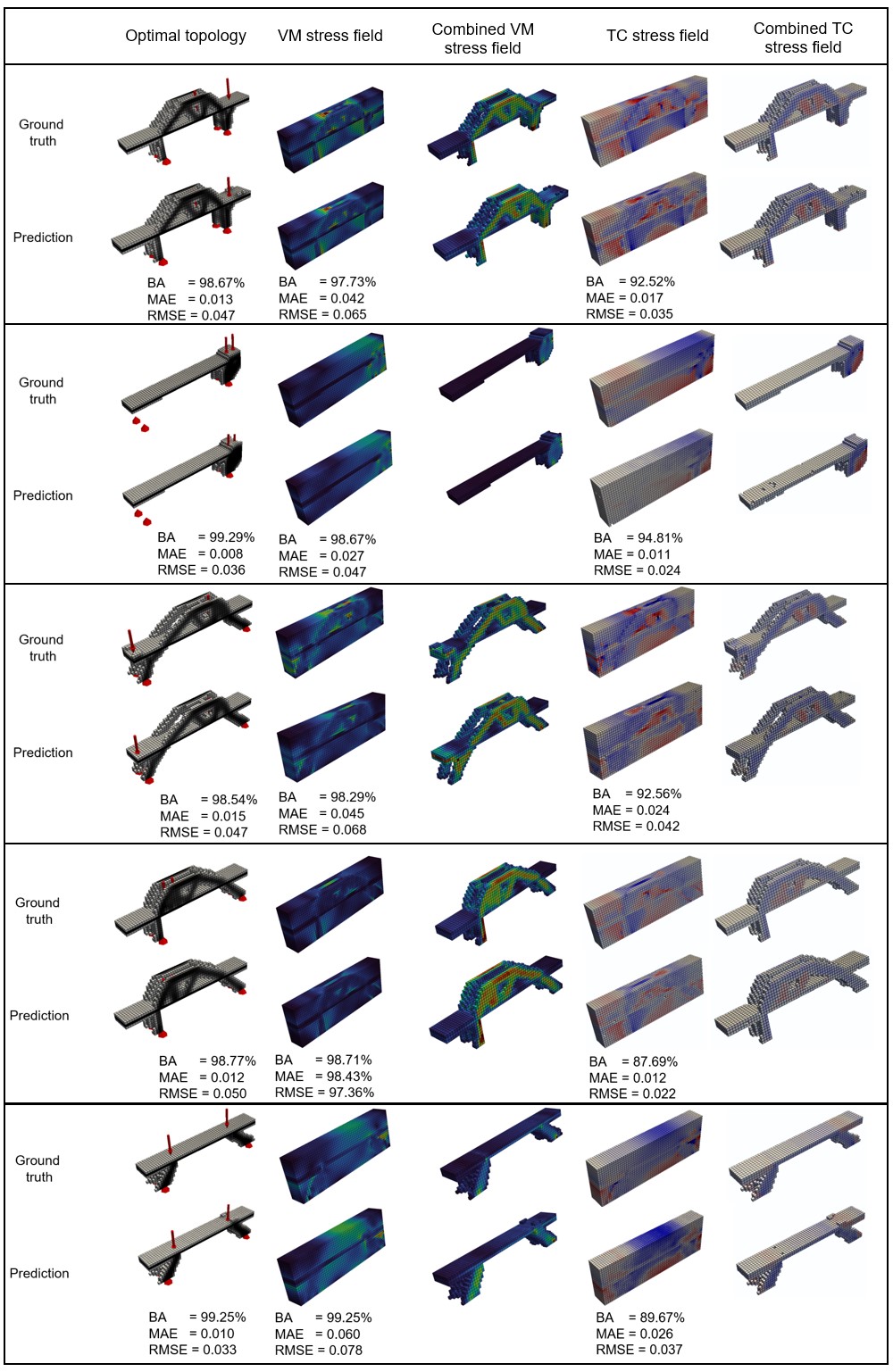}
    \caption{3D bridge. Exemplary comparisons of optimal topologies, VM and TC stress fields, and corresponding combined fields, obtained through the SIMP-based procedure and the proposed framework. Results in each box are obtained for varying positions of the acting forces and of the supports. For each case, the accuracy of the predicted topology is measured in terms of binary accuracy, mean absolute error and root mean squared error.
    \label{fig:3DBridge_results_VM_TC}}
\end{figure}

Some exemplary results, comprehensive of the predictions for the $\boldsymbol{x}_\mathtt{VM}$ and $\boldsymbol{x}_\mathtt{TC}$ stress fields, as well as of the relative combined fields, are reported in \fig\ref{fig:3DBridge_results_VM_TC}. For each test case, the top row shows the ground truth results, generated using the SIMP-based algorithm, and the bottom row reports the corresponding approximations, predicted through the proposed DL model. An overall assessment of the predictive capabilities of the proposed framework is shown in \tab\ref{tab:3DBridge_accuracy}, which reports the
BA, MAE, and RMSE performance indicators for the predicted $\boldsymbol{x}$, $\boldsymbol{x}_\mathtt{VM}$ and $\boldsymbol{x}_\mathtt{TC}$ fields, averaged over the 500 testing instances.

\begin{table}[h!]
\caption{3D bridge. Accuracy of the proposed DL model against the ground truth SIMP-based procedure. Results are reported in terms of binary accuracy, mean absolute error and root mean squared error, all averaged over $500$ testing instances.}
    \centering
    \begin{tabular}{cccc}
    \toprule
    \mbox{Accuracy metrics} & \mbox{Optimal topologies} &  \mbox{$\boldsymbol{x}_\mathtt{VM}$ stress field } & \mbox{$\boldsymbol{x}_\mathtt{TC}$ stress field }\\
    \toprule
    \mbox{BA}               & \mbox{98.50\%}    & \mbox{98.27\%}        & \mbox{89.96\%}\\
    \mbox{MAE}              & \mbox{0.015}      & \mbox{0.037}          & \mbox{0.021}\\
    \mbox{RMSE}              & \mbox{0.051}      & \mbox{0.057}          & \mbox{0.038}\\
    \bottomrule
    \end{tabular}
    \label{tab:3DBridge_accuracy}
\end{table}

\paragraph{Discussion of the results.}
For this second case study, the results obtained for the prediction of optimal topologies reach improved accuracy values compared to the 2D MBB beam case, yielding $98.50\%$ for the BA, $0.015$ for the MAE, and $0.51$ for the RMSE, respectively. The $\boldsymbol{x}_\mathtt{VM}$ predictions score $98.27\%$, $0.037$, and $0.057$, while the $\boldsymbol{x}_\mathtt{TC}$ predictions score $89.96\%$, $0.021$, and $0.038$, respectively, for the same accuracy metrics. In the following, we discuss some aspects behind the obtained results.

\begin{figure}[h!]
\centering
\captionsetup[subfigure]{justification=centering}\subfloat[\label{fig:hist_beam}]{\includegraphics[width=.4\textwidth]{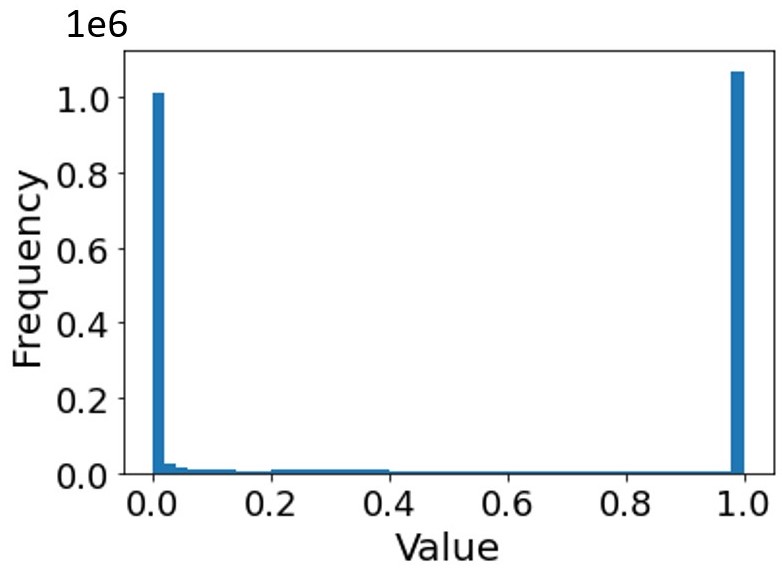}}\hspace{0.5cm}
\subfloat[\label{fig:hist_bridge}]{\includegraphics[width=.4\textwidth]{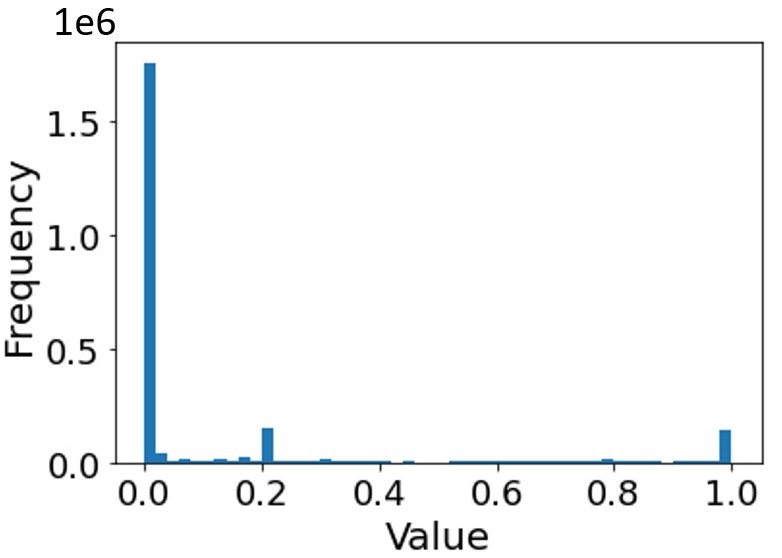}}
\caption{Distribution of the density values characterizing the testing optimal topologies for (a) the 2D MBB beam case and (b) the 3D bridge case. The frequency refers to the number of elements for which a certain density value is reported over the testing instances.\label{fig:density_hist}}
\end{figure}

\begin{itemize}
    \item The prediction of optimal topologies shows a significant improvement, with MAE and RMSE values more than halved compared to the 2D MBB beam case. This is likely because the target optimal topologies for varying input parameters exhibit smaller variability, which typically results in a variety of similar structural forms. On the other hand, the 2D MBB beam topologies feature numerous highly variable and complex elements, which hinder the accurate learning of the corresponding feature maps by the DL models. 
    \item A second aspect worth mentioning is the presence of solid regions for the bridge deck and of the passive domain for the empty space under the deck. These regions characterize every topology, collectively covering $30\%$ of the design domain. This recurring pattern facilitates accurate learning and generalization by the DL models to unseen test cases.  
    \item Another reason behind the improved values of MAE and RMSE performance indicators for the density field predictions is the different distribution characterizing the density field data in the two case studies. As shown in the bar charts in \fig\ref{fig:density_hist}, the optimal topologies for the 2D MBB beam exhibit a markedly $0-1$ binary distribution, while those for the 3D bridge show a more evenly spread range of density values. In the latter case, an incorrect prediction is more likely to be close to the correct value, thus not significantly impacting the accuracy metric.
\end{itemize}

\section{Conclusions}
\label{sec:conclusions}

This paper has presented a multi-stage deep learning strategy to efficiently predict optimal topologies in both two and three-dimensional design tasks. The proposed strategy is inspired by a deep learning-based dimensionality reduction procedure \cite{Fresca2020} for reduced-order modeling of parametrized differential problems. Real-time prediction of optimal topologies is achieved by leveraging a fully-connected model to map the parametric input space onto a low-dimensional latent space, which feeds the decoder branch of an autoencoder model to approximate the desired optimal topology.

One of the main advantages of the proposed method is the capability of capturing common patterns in the training dataset, enabling efficient data compression without significant loss of information. This allows for learning an effective functional link between the parametric input space and the low-dimensional latent space through the fully-connected model, which is general enough to be applied to both 2D and 3D problems. Moreover, the method is capable of predicting physics-based information about the problem, such as the fields of Von Mises stress and dominant principal stress, alongside the optimal topology field, without the need for physics solvers to be included in the prediction stage.

Once trained, the proposed framework enables the almost instantaneous prediction of the three aforementioned fields.  With reference to the topology optimization of a 3D bridge case study, the predictions of the optimal density field has yielded mean absolute error and root mean squared error vales of $0.015$ and $0.051$, respectively, with an average run-time of 0.13 seconds only. This provides clear advantages over traditional and computationally expensive topology optimization algorithms, which can be particularly useful during the conceptual design phase of a project, where the quick and accurate exploration of feasible solutions is crucial.

The future activities will be devoted to explore test cases with larger design domains, to show a more real-world applicability of the proposed method. Another aspect of interest concerns the use of more principled sampling strategies to populate the training dataset, potentially allowing either for higher accuracy scores or smaller training sets yielding comparable accuracy scores. Finally, the considered topology optimization framework could also be extended to include other types of constraints, such as stress constraints.

\section*{Online application availability}

To showcase the real-time capabilities of the proposed topology optimization framework, an online application is made available at \cite{app}. The app features a user-friendly graphical interface, allowing to vary the loading and boundary conditions for the 3D bridge case and dynamically update the predicted topology, which is displayed in real-time. The app is implemented in \texttt{JavaScript}, \texttt{HTML} and \texttt{CSS}, and makes use of the open-source \texttt{Three.js} 3D visualization library \cite{threejs}.

\vspace{14pt} 
\noindent{\bf Data Availability Statement.}
The data that support the findings of this study are available from the corresponding author upon reasonable request.

\bibliographystyle{ieeetr}
\bibliographystyle{ieeetr}

\appendix
\newpage
\section{Implementation details}
\label{sec:implementation}

In this Appendix, we discuss the implementation details of the DL models discussed in \sez\ref{sec:ML_formulation}. The architectures, as well as the relevant hyperparameters and training options, have been chosen through a preliminary study, aimed at minimizing $\mathcal{L}_\mathtt{AE}$ and $\mathcal{L}_\mathtt{FC}$, while retaining the generalization capabilities of $\mathcal{NN}_\mathtt{ENC}$, $\mathcal{NN}_\mathtt{DEC}$, and $\mathcal{NN}_\mathtt{FC}$.

$\mathcal{NN}_\mathtt{ENC}$ and $\mathcal{NN}_\mathtt{DEC}$ are set as the encoder and decoder of a CNN-AE model, whose architecture is described in \tab\ref{tab:NN_2D_AE_arch} and in \tab\ref{tab:NN_3D_AE_arch} for the 2D and 3D cases, respectively. The encoding branch consists of a stack of four convolutional units, each one featuring two convolutional layers, followed by a max pooling layer and a batch normalization layer. The output is then flattened and run through a fully-connected layer to provide a low-dimensional feature space featuring $N_L=40$ neurons. This bottleneck layer is linked to the decoding branch by means of a fully-connected layer, whose output is then reshaped before undergo through a stack of four transposed convolutional units in order to reconstruct the input image. Each transposed convolutional unit in the decoder features two convolutional layers, a transposed convolutional layer and a batch normalization layer. A cropping layer is employed at the end of the 3D decoder, to address inconsistencies in the output shapes due to rounding. All convolutional layers feature $3\times3$ or $3\times3\times3$ kernels for the 2D and 3D cases, respectively. These layers are all ReLU-activated, except the last one that is sigmoid-activated in the case of the density and VM stress fields, and has no activation function in the case of the TC stress field. No activation is applied to the bottleneck layer, while the other fully-connected layer is ReLU-activated. 

Using the Xavier's weight initialization \cite{art:Glorot}, the loss function $\mathcal{L}_\mathtt{AE}$ is minimized using the first-order stochastic gradient descent optimizer Adam \cite{art:Adam}, for a maximum of $250$ allowed epochs. The learning rate $\eta_\mathtt{AE}$ is initially set to $10^{-4}$, and decreased with a linear decay schedule. The optimization is carried out considering an $80:20$ splitting ratio for training and validation purposes, so that $20\%$ of the data is randomly taken and set aside to monitor the learning process. In particular, we use an early stopping strategy to interrupt learning, whenever the loss function value attained on the validation set does not decrease for a prescribed number of patience epochs in a row. The hyperparameters and training options for $\mathcal{NN}_\mathtt{2D,AE}$ and for $\mathcal{NN}_\mathtt{3D,AE}$ are reported in \tab\ref{tab:NN_2D_AE_hyper} and in \tab\ref{tab:NN_3D_AE_hyper}, respectively. 

$\mathcal{NN}_\mathtt{FC}$ consists of three fully-connected layers featuring $720, 720$ and $N_L=40$ neurons, respectively. Each pair of layers is separated by a batch normalization layer.  No activation is applied to the output layer, while the hidden layers feature a ReLU activation function. The architecture of $\mathcal{NN}_\mathtt{FC}$ is outlined in \tab\ref{tab:NN_FCM_arch}, where the only difference between the 2D and 3D cases is the size $\mathtt{N_{par}}$ of the input layer, which is problem specific. Also in this case, the optimization is carried out using Adam together with the Xavier's weight initialization. The learning rate $\eta_\mathtt{FC}$ is decreased as the training advances using a linear decay schedule, and an early stop strategy is used to prevent overfitting by considering an $80:20$ splitting ratio for training and validation purposes. The relevant hyperparameters and the training options are summarized in \tab\ref{tab:NN_FCM_hyper}.

\begin{table}
\caption{$\mathcal{NN}_\mathtt{2D,ENC}, \mathcal{NN}_\mathtt{2D,DEC}$ - (a) employed architecture, and (b) selected hyperparameters and training options.}
      \centering
       \subfloat[\label{tab:NN_2D_AE_arch}]{
       \scriptsize
\begin{tabular}{llll}
    \toprule
    \mbox{Layer} & \mbox{Output shape} &  \mbox{Activation} & \mbox{Input}\\
    \toprule
    \mbox{0 - Input}                & \mbox{$(B_\mathtt{AE},120, 40, 1)$}     & \mbox{None}               & \mbox{--}\\
    \mbox{1 - Conv2D}               & \mbox{$(B_\mathtt{AE},120, 40, 128)$}   & \mbox{$\mathtt{ReLU}$}    & \mbox{0}\\
    \mbox{2 - Conv2D}               & \mbox{$(B_\mathtt{AE},120, 40, 128)$}   & \mbox{$\mathtt{ReLU}$}    & \mbox{1}\\
    \mbox{3 - MaxPooling2D}         & \mbox{$(B_\mathtt{AE},60, 20, 128)$}    & \mbox{None}               & \mbox{2}\\
    \mbox{4 - BatchNormalization}   & \mbox{$(B_\mathtt{AE},60, 20, 128)$}    & \mbox{None}               & \mbox{3}\\
    \mbox{5 - Conv2D}               & \mbox{$(B_\mathtt{AE},60, 20, 64)$}     & \mbox{$\mathtt{ReLU}$}    & \mbox{4}\\
    \mbox{6 - Conv2D}               & \mbox{$(B_\mathtt{AE},60, 20, 64)$}     & \mbox{$\mathtt{ReLU}$}    & \mbox{5}\\
    \mbox{7 - MaxPooling2D}         & \mbox{$(B_\mathtt{AE},30, 10, 64)$}     & \mbox{None}               & \mbox{6}\\
    \mbox{8 - BatchNormalization}   & \mbox{$(B_\mathtt{AE},30, 10, 64)$}     & \mbox{None}               & \mbox{7}\\  
    \mbox{9 - Conv2D}               & \mbox{$(B_\mathtt{AE},30, 10, 32)$}     & \mbox{$\mathtt{ReLU}$}    & \mbox{8}\\
    \mbox{10 - Conv2D}              & \mbox{$(B_\mathtt{AE},30, 10, 32)$}     & \mbox{$\mathtt{ReLU}$}    & \mbox{9}\\
    \mbox{11 - MaxPooling2D}        & \mbox{$(B_\mathtt{AE},15, 5, 32)$}      & \mbox{None}               & \mbox{10}\\
    \mbox{12 - BatchNormalization}  & \mbox{$(B_\mathtt{AE},15, 5, 32)$}      & \mbox{None}               & \mbox{11}\\   
    \mbox{13 - Conv2D}              & \mbox{$(B_\mathtt{AE},15, 5, 32)$}      & \mbox{$\mathtt{ReLU}$}    & \mbox{12}\\
    \mbox{14 - Conv2D}              & \mbox{$(B_\mathtt{AE},15, 5, 32)$}      & \mbox{$\mathtt{ReLU}$}    & \mbox{13}\\
    \mbox{15 - Reshape}             & \mbox{$(B_\mathtt{AE},2400)$}           & \mbox{$\mathtt{ReLU}$}    & \mbox{14}\\
    \mbox{16 - Dense}               & \mbox{$(B_\mathtt{AE},N_L=40)$}             & \mbox{None} & \mbox{15}\\
    
    \mbox{17 - Dense}               & \mbox{$(B_\mathtt{AE},2400)$}           & \mbox{$\mathtt{ReLU}$}   & \mbox{16}\\
    \mbox{18 - Reshape}             & \mbox{$(B_\mathtt{AE},15, 5, 32)$}      & \mbox{None}               & \mbox{17}\\
    \mbox{19 - BatchNormalization}  & \mbox{$(B_\mathtt{AE},15, 5, 32)$}      & \mbox{None}               & \mbox{18}\\ 
    \mbox{20 - Conv2D}              & \mbox{$(B_\mathtt{AE},15, 5, 32)$}      & \mbox{$\mathtt{ReLU}$}    & \mbox{19}\\
    \mbox{21 - Conv2D}              & \mbox{$(B_\mathtt{AE},15, 5, 32)$}      & \mbox{$\mathtt{ReLU}$}    & \mbox{20}\\
    \mbox{22 - Conv2DTranspose}     & \mbox{$(B_\mathtt{AE},30, 10, 32)$}     & \mbox{$\mathtt{ReLU}$}    & \mbox{21}\\
    \mbox{23 - BatchNormalization}  & \mbox{$(B_\mathtt{AE},30, 10, 32)$}     & \mbox{None}               & \mbox{22}\\ 
    \mbox{24 - Conv2D}              & \mbox{$(B_\mathtt{AE},30, 10, 32)$}     & \mbox{$\mathtt{ReLU}$}    & \mbox{23}\\
    \mbox{25 - Conv2D}              & \mbox{$(B_\mathtt{AE},30, 10, 32)$}     & \mbox{$\mathtt{ReLU}$}    & \mbox{24}\\
    \mbox{26 - Conv2DTranspose}     & \mbox{$(B_\mathtt{AE},60, 20, 64)$}     & \mbox{$\mathtt{ReLU}$}    & \mbox{25}\\
    \mbox{27 - BatchNormalization}  & \mbox{$(B_\mathtt{AE},60, 20, 64)$}     & \mbox{None}               & \mbox{26}\\ 
    \mbox{28 - Conv2D}              & \mbox{$(B_\mathtt{AE},60, 20, 64)$}     & \mbox{$\mathtt{ReLU}$}    & \mbox{27}\\
    \mbox{29 - Conv2D}              & \mbox{$(B_\mathtt{AE},60, 20, 64)$}     & \mbox{$\mathtt{ReLU}$}    & \mbox{28}\\
    \mbox{30 - Conv2DTranspose}     & \mbox{$(B_\mathtt{AE},120, 40, 128)$}   & \mbox{$\mathtt{ReLU}$}    & \mbox{29}\\
    \mbox{31 - BatchNormalization}  & \mbox{$(B_\mathtt{AE},120, 40, 128)$}   & \mbox{None}               & \mbox{30}\\ 
    \mbox{32 - Conv2D}              & \mbox{$(B_\mathtt{AE},120, 40, 128)$}   & \mbox{$\mathtt{ReLU}$}    & \mbox{31}\\
    \mbox{33 - Conv2D}              & \mbox{$(B_\mathtt{AE},120, 40, 128)$}   & \mbox{$\mathtt{ReLU}$}    & \mbox{32}\\    
    \mbox{34 - Conv2D}              & \mbox{$(B_\mathtt{AE},120, 40, 1)$}     & \mbox{$\mathtt{Sigmoid}$/None}    & \mbox{33}\\
        \bottomrule
          \end{tabular}
          }
       \subfloat[\label{tab:NN_2D_AE_hyper}]{
       \scriptsize
  \begin{tabular}{ll}
    \toprule
        \mbox{Convolution kernel size:}         & \mbox{$3\times3$}\\
        \mbox{Weight initializer:}              & \mbox{Xavier}\\
        \mbox{Optimizer:}                       & \mbox{Adam}\\
        \mbox{Batch size:}                      & \mbox{$B_\mathtt{AE}=32$}\\
        \mbox{Initial learning rate:}           & \mbox{$\eta_\mathtt{AE}=10^{-4}$}\\
        \mbox{Allowed epochs:}                  & \mbox{$250$}\\
        \mbox{Learning schedule:}               & \mbox{Linear}\\
        \mbox{Early stop patience:}             & \mbox{30 epochs}\\
        \mbox{Train-val split:}                 & \mbox{$80:20$}\\
    \bottomrule
  \end{tabular}
  }
\end{table}

\begin{table}
\caption{$\mathcal{NN}_\mathtt{3D,ENC}, \mathcal{NN}_\mathtt{3D,DEC}$ - (a) employed architecture, and (b) selected hyperparameters and training options.}
      \centering
       \subfloat[\label{tab:NN_3D_AE_arch}]{
       \scriptsize
\begin{tabular}{llll}
    \toprule
    \mbox{Layer} & \mbox{Output shape} &  \mbox{Activation} & \mbox{Input}\\
    \toprule
    \mbox{0 - Input}                & \mbox{$(B_\mathtt{AE},60, 20, 4, 1)$}       & \mbox{None}               & \mbox{--}\\
    \mbox{1 - Conv3D}               & \mbox{$(B_\mathtt{AE},60, 20, 4, 128)$}     & \mbox{$\mathtt{ReLU}$}    & \mbox{0}\\
    \mbox{2 - Conv3D}               & \mbox{$(B_\mathtt{AE},60, 20, 4, 128)$}     & \mbox{$\mathtt{ReLU}$}    & \mbox{1}\\
    \mbox{3 - MaxPooling3D}         & \mbox{$(B_\mathtt{AE},30, 10, 2, 128)$}     & \mbox{None}               & \mbox{2}\\
    \mbox{4 - BatchNormalization}   & \mbox{$(B_\mathtt{AE},30, 10, 2, 128)$}     & \mbox{None}               & \mbox{3}\\
    \mbox{5 - Conv3D}               & \mbox{$(B_\mathtt{AE},30, 10, 2, 64)$}      & \mbox{$\mathtt{ReLU}$}    & \mbox{4}\\
    \mbox{6 - Conv3D}               & \mbox{$(B_\mathtt{AE},30, 10, 2, 64)$}      & \mbox{$\mathtt{ReLU}$}    & \mbox{5}\\
    \mbox{7 - MaxPooling3D}         & \mbox{$(B_\mathtt{AE},15, 5, 1, 64)$}       & \mbox{None}               & \mbox{6}\\
    \mbox{8 - BatchNormalization}   & \mbox{$(B_\mathtt{AE},15, 5, 1, 64)$}       & \mbox{None}               & \mbox{7}\\  
    \mbox{9 - Conv3D}               & \mbox{$(B_\mathtt{AE},15, 5, 1, 32)$}       & \mbox{$\mathtt{ReLU}$}    & \mbox{8}\\
    \mbox{10 - Conv3D}              & \mbox{$(B_\mathtt{AE},15, 5, 1, 32)$}       & \mbox{$\mathtt{ReLU}$}    & \mbox{9}\\
    \mbox{11 - MaxPooling3D}        & \mbox{$(B_\mathtt{AE},8, 3, 1, 32)$}        & \mbox{None}               & \mbox{10}\\
    \mbox{12 - BatchNormalization}  & \mbox{$(B_\mathtt{AE},8, 3, 1, 32)$}        & \mbox{None}               & \mbox{11}\\   
    \mbox{13 - Conv3D}              & \mbox{$(B_\mathtt{AE},8, 3, 1, 32)$}        & \mbox{$\mathtt{ReLU}$}    & \mbox{12}\\
    \mbox{14 - Conv3D}              & \mbox{$(B_\mathtt{AE},8, 3, 1, 32)$}        & \mbox{$\mathtt{ReLU}$}    & \mbox{13}\\
    \mbox{15 - Reshape}             & \mbox{$(B_\mathtt{AE},768)$}                & \mbox{$\mathtt{ReLU}$}    & \mbox{14}\\
    \mbox{16 - Dense}               & \mbox{$(B_\mathtt{AE},N_L=40)$}                 & \mbox{None} & \mbox{15}\\
    
    \mbox{17 - Dense}               & \mbox{$(B_\mathtt{AE},768)$}                & \mbox{$\mathtt{ReLU}$}   & \mbox{16}\\
    \mbox{18 - Reshape}             & \mbox{$(B_\mathtt{AE},8, 3, 1, 32)$}        & \mbox{None}               & \mbox{17}\\
    \mbox{19 - BatchNormalization}  & \mbox{$(B_\mathtt{AE},8, 3, 1, 32)$}        & \mbox{None}               & \mbox{18}\\ 
    \mbox{20 - Conv3D}              & \mbox{$(B_\mathtt{AE},8, 3, 1, 32)$}        & \mbox{$\mathtt{ReLU}$}    & \mbox{19}\\
    \mbox{21 - Conv3D}              & \mbox{$(B_\mathtt{AE},8, 3, 1, 32)$}        & \mbox{$\mathtt{ReLU}$}    & \mbox{20}\\
    \mbox{22 - Conv3DTranspose}     & \mbox{$(B_\mathtt{AE},16, 6, 2, 32)$}       & \mbox{$\mathtt{ReLU}$}    & \mbox{21}\\
    \mbox{23 - BatchNormalization}  & \mbox{$(B_\mathtt{AE},16, 6, 2, 32)$}       & \mbox{None}               & \mbox{22}\\ 
    \mbox{24 - Conv3D}              & \mbox{$(B_\mathtt{AE},16, 6, 2, 32)$}       & \mbox{$\mathtt{ReLU}$}    & \mbox{23}\\
    \mbox{25 - Conv3D}              & \mbox{$(B_\mathtt{AE},16, 6, 2, 32)$}       & \mbox{$\mathtt{ReLU}$}    & \mbox{24}\\
    \mbox{26 - Conv3DTranspose}     & \mbox{$(B_\mathtt{AE},32, 12, 4, 64)$}      & \mbox{$\mathtt{ReLU}$}    & \mbox{25}\\
    \mbox{27 - BatchNormalization}  & \mbox{$(B_\mathtt{AE},32, 12, 4, 64)$}      & \mbox{None}               & \mbox{26}\\ 
    \mbox{28 - Conv3D}              & \mbox{$(B_\mathtt{AE},32, 12, 4, 64)$}      & \mbox{$\mathtt{ReLU}$}    & \mbox{27}\\
    \mbox{29 - Conv3D}              & \mbox{$(B_\mathtt{AE},32, 12, 4, 64)$}      & \mbox{$\mathtt{ReLU}$}    & \mbox{28}\\
    \mbox{30 - Conv3DTranspose}     & \mbox{$(B_\mathtt{AE},64, 24, 8, 128)$}     & \mbox{$\mathtt{ReLU}$}    & \mbox{29}\\
    \mbox{31 - BatchNormalization}  & \mbox{$(B_\mathtt{AE},64, 24, 8, 128)$}     & \mbox{None}               & \mbox{30}\\ 
    \mbox{32 - Conv3D}              & \mbox{$(B_\mathtt{AE},64, 24, 8, 128)$}     & \mbox{$\mathtt{ReLU}$}    & \mbox{31}\\
    \mbox{33 - Conv3D}              & \mbox{$(B_\mathtt{AE},64, 24, 8, 128)$}     & \mbox{$\mathtt{ReLU}$}    & \mbox{32}\\    
    \mbox{34 - Conv3D}              & \mbox{$(B_\mathtt{AE},64, 24, 8, 1)$}       & \mbox{$\mathtt{Sigmoid}$/None}    & \mbox{33}\\    
    \mbox{35 - Cropping3D}          & \mbox{$(B_\mathtt{AE},60, 20, 4, 1)$}       & \mbox{None}               & \mbox{34}\\
        \bottomrule
          \end{tabular}
          }
       \subfloat[\label{tab:NN_3D_AE_hyper}]{
       \scriptsize
  \begin{tabular}{ll}
    \toprule
        \mbox{Convolution kernel size:}         & \mbox{$3\times3\times3$}\\
        \mbox{Weight initializer:}              & \mbox{Xavier}\\
        \mbox{Optimizer:}                       & \mbox{Adam}\\
        \mbox{Batch size:}                      & \mbox{$B_\mathtt{AE}=32$}\\
        \mbox{Initial learning rate:}           & \mbox{$\eta_\mathtt{AE}=10^{-4}$}\\
        \mbox{Allowed epochs:}                  & \mbox{$250$}\\
        \mbox{Learning schedule:}               & \mbox{Linear}\\
        \mbox{Early stop patience:}             & \mbox{30 epochs}\\
        \mbox{Train-val split:}                 & \mbox{$80:20$}\\
    \bottomrule
  \end{tabular}
  }
\end{table}

\begin{table}
\caption{$\mathcal{NN}_\mathtt{FC}$ - (a) employed architecture, and (b) selected hyperparameters and training options.}
      \centering
       \subfloat[\label{tab:NN_FCM_arch}]{
       \scriptsize
\begin{tabular}{llll}
    \toprule
    \mbox{Layer} & \mbox{Output shape} &  \mbox{Activation} & \mbox{Input}\\
    \toprule
    \mbox{0 - Input}                & \mbox{$(B_\mathtt{FC},\mathtt{N_{par}})$}      & \mbox{None}               & \mbox{--}\\
    \mbox{1 - Dense}                & \mbox{$(B_\mathtt{FC},720)$}    & \mbox{$\mathtt{ReLU}$}    & \mbox{0}\\
    \mbox{2 - BatchNormalization}   & \mbox{$(B_\mathtt{FC},720)$}    & \mbox{None}               & \mbox{1}\\
    \mbox{3 - Dense}                & \mbox{$(B_\mathtt{FC},720)$}    & \mbox{$\mathtt{ReLU}$}    & \mbox{2}\\
    \mbox{4 - BatchNormalization}   & \mbox{$(B_\mathtt{FC},720)$}    & \mbox{None}               & \mbox{3}\\  
    \mbox{5 - Dense}                & \mbox{$(B_\mathtt{FC},N_L=40)$}     & \mbox{None}    & \mbox{4}\\
        \bottomrule
          \end{tabular}
          }
       \subfloat[\label{tab:NN_FCM_hyper}]{
       \scriptsize
  \begin{tabular}{ll}
    \toprule
        \mbox{Weight initializer:}              & \mbox{Xavier}\\
        \mbox{Optimizer:}                       & \mbox{Adam}\\
        \mbox{Batch size:}                      & \mbox{$B_\mathtt{FC}=32$}\\
        \mbox{Initial learning rate:}           & \mbox{$\eta_\mathtt{FC}=10^{-4}$}\\
        \mbox{Allowed epochs:}                  & \mbox{$400$}\\
        \mbox{Learning schedule:}               & \mbox{Linear}\\
        \mbox{Early stop patience:}             & \mbox{30 epochs}\\
        \mbox{Train-val split:}                 & \mbox{$80:20$}\\
    \bottomrule
  \end{tabular}
  }
\end{table}

\newpage

\section{Comparative study for varying architectures}
\label{sec:comparative_exp}

In this appendix, we investigate the impact of specific choices about the architecture of DL models on the accuracy achieved in approximating optimal topologies for the 2D MBB beam case study. Initially, we train new autoencoders and the fully-connected models with varying numbers of layers. Subsequently, we test each possible combination for the combined use of the fully-connected model and the decoding branch of the autoencoder. The outcomes of these additional experiments are then compared with the results obtained using the reference architectures described in Appendix~\ref{sec:implementation}.

In the reference autoencoder architecture $\mathcal{NN}_\mathtt{AE}$, the encoding branch comprises a stack of four convolutional units, each featuring two convolutional layers, followed by a max pooling layer and a batch normalization layer. The decoding branch consists of a stack of four transposed convolutional units, each featuring two convolutional layers, a transposed convolutional layer and a batch normalization layer. The reference fully-connected model architecture $\mathcal{NN}_\mathtt{FC}$ consists of three fully-connected layers, with each pair of layers separated by a batch normalization layer.

We train two additional autoencoders and two additional fully connected models. The first new autoencoder, denoted as $\mathcal{NN}_\mathtt{AE}^+$, features an additional convolutional unit in the encoding branch and an extra transposed convolutional unit in the decoding branch. The second one, denoted as $\mathcal{NN}_\mathtt{AE}^-$, features one less convolutional unit in the encoding branch and one fewer transposed convolutional unit in the decoding branch compared to the reference architecture. Similarly, the two new fully-connected models, namely $\mathcal{NN}_\mathtt{FC}^+$ and $\mathcal{NN}_\mathtt{FC}^-$, are obtained by adding and removing a fully-connected layer from the reference architecture. Considering all the possible combinations of fully-connected models and decoding branches, we obtain eight new alternative prediction models.
 
An overall quantitative assessment of the predictive capabilities of the new models is presented in \tab\ref{tab:2DMBB_accuracy_combinations}, which reports the BA, MAE, and RMSE performance indicators for the predicted density field. These values are averaged over $500$ testing instances. The first column refers to the performance achieved by the reference model, as already as shown in \tab\ref{tab:2DMBB_accuracy}, and serves as the control group against which the performance of other prediction models can be compared. The relative difference in the performance of the alternative models compared to the reference one is reported in \tab\ref{tab:2DMBB_accuracy_combinations_percentages}, in terms of percentage variations averaged over 500 testing instances. These results demonstrate that adding an extra dense layer in the fully-connected model results in improvements across all accuracy metric, regardless of the decoding branch employed to reconstruct the desired optimal topology from its latent representation. This suggests that the enhancements arising from improved representation capabilities for the fully-connected model are more significant than those resulting from modifications to the architecture of the autoencoder. This observation is further supported by the case in which $\mathcal{NN}_\mathtt{AE}^+$ and $\mathcal{NN}_\mathtt{FC}^-$ are considered as source models, leading to a deterioration in performance across all accuracy metrics.

\begin{table}[h!]
\small
\caption{MBB beam. Accuracy of the proposed DL model against the ground truth SIMP-based procedure for varying architectures of the autoencoder and the fully-connected model. Results are reported in terms of binary accuracy, mean absolute error and root mean squared error, all averaged over $500$ testing instances.}
    \centering
    \begin{tabular}{cccccccccc}
    \toprule
    \multicolumn{1}{p{1.2cm}}{\centering\mbox{Source}\\\mbox{models}} & 
    \multicolumn{1}{p{1.1cm}}{\centering \mbox{$\mathcal{NN}_\mathtt{AE}$} \\ \mbox{$\mathcal{NN}_\mathtt{FC}$}} &
    \multicolumn{1}{p{1.1cm}}{\centering \mbox{$\mathcal{NN}_\mathtt{AE}$} \\ \mbox{$\mathcal{NN}_\mathtt{FC}^+$}} &
    \multicolumn{1}{p{1.1cm}}{\centering \mbox{$\mathcal{NN}_\mathtt{AE}$} \\ \mbox{$\mathcal{NN}_\mathtt{FC}^-$}} &
    \multicolumn{1}{p{1.1cm}}{\centering \mbox{$\mathcal{NN}_\mathtt{AE}^+$} \\ \mbox{$\mathcal{NN}_\mathtt{FC}$}} &
    \multicolumn{1}{p{1.1cm}}{\centering \mbox{$\mathcal{NN}_\mathtt{AE}^+$} \\ \mbox{$\mathcal{NN}_\mathtt{FC}^+$}} &
    \multicolumn{1}{p{1.1cm}}{\centering \mbox{$\mathcal{NN}_\mathtt{AE}^+$} \\ \mbox{$\mathcal{NN}_\mathtt{FC}^-$}} &
    \multicolumn{1}{p{1.1cm}}{\centering \mbox{$\mathcal{NN}_\mathtt{AE}^-$} \\ \mbox{$\mathcal{NN}_\mathtt{FC}$}} &
    \multicolumn{1}{p{1.1cm}}{\centering \mbox{$\mathcal{NN}_\mathtt{AE}^-$} \\ \mbox{$\mathcal{NN}_\mathtt{FC}^+$}} &
    \multicolumn{1}{p{1.1cm}}{\centering \mbox{$\mathcal{NN}_\mathtt{AE}^-$} \\ \mbox{$\mathcal{NN}_\mathtt{FC}^-$}} \\
    \toprule
    \mbox{BA} & \mbox{$96.46\%$}  & \mbox{$97.02\%$} & \mbox{$93.76\%$} & \mbox{$96.77\%$} & \mbox{$97.01\%$} & \mbox{$94.66\%$} & \mbox{$96.27\%$} & \mbox{$96.82\%$} & \mbox{$92.10\%$} \\
    
    \mbox{MAE} & \mbox{$0.035$} & \mbox{$0.030$} & \mbox{$0.062$} & \mbox{$0.032$} & \mbox{$0.030$} & \mbox{$0.053$} 
     & \mbox{$0.037$} & \mbox{$0.032$} & \mbox{$0.078$} \\
    
    \mbox{RMSE} & \mbox{$0.107$}  & \mbox{$0.093$}  & \mbox{$0.168$}  & \mbox{$0.098$}  & \mbox{$0.094$}  & \mbox{$0.144$} & \mbox{$0.110$}  & \mbox{$0.098$}  & \mbox{$0.198$} \\
    \bottomrule
    \end{tabular}
    
    \label{tab:2DMBB_accuracy_combinations}
\end{table}

\begin{table}[h!]
\small
\caption{MBB beam. Relative difference in the accuracy attained by the alternative models with respect to the reference one. Results are reported in terms of percentage variation in binary accuracy, mean absolute error and root mean squared error, all averaged over $500$ testing instances.}
    \centering
    \begin{tabular}{ccccccccc}
    \toprule
    \multicolumn{1}{p{1.2cm}}{\centering\mbox{Source}\\\mbox{models}} & 
    \multicolumn{1}{p{1.1cm}}{\centering \mbox{$\mathcal{NN}_\mathtt{AE}$} \\ \mbox{$\mathcal{NN}_\mathtt{FC}^+$}} &
    \multicolumn{1}{p{1.1cm}}{\centering \mbox{$\mathcal{NN}_\mathtt{AE}$} \\ \mbox{$\mathcal{NN}_\mathtt{FC}^-$}} &
    \multicolumn{1}{p{1.1cm}}{\centering \mbox{$\mathcal{NN}_\mathtt{AE}^+$} \\ \mbox{$\mathcal{NN}_\mathtt{FC}$}} &
    \multicolumn{1}{p{1.1cm}}{\centering \mbox{$\mathcal{NN}_\mathtt{AE}^+$} \\ \mbox{$\mathcal{NN}_\mathtt{FC}^+$}} &
    \multicolumn{1}{p{1.1cm}}{\centering \mbox{$\mathcal{NN}_\mathtt{AE}^+$} \\ \mbox{$\mathcal{NN}_\mathtt{FC}^-$}} &
    \multicolumn{1}{p{1.1cm}}{\centering \mbox{$\mathcal{NN}_\mathtt{AE}^-$} \\ \mbox{$\mathcal{NN}_\mathtt{FC}$}} &
    \multicolumn{1}{p{1.1cm}}{\centering \mbox{$\mathcal{NN}_\mathtt{AE}^-$} \\ \mbox{$\mathcal{NN}_\mathtt{FC}^+$}} &
    \multicolumn{1}{p{1.1cm}}{\centering \mbox{$\mathcal{NN}_\mathtt{AE}^-$} \\ \mbox{$\mathcal{NN}_\mathtt{FC}^-$}} \\
    \toprule
    \mbox{BA} & \mbox{$+0.6\%$} & \mbox{$-2.8\%$} & \mbox{$+0.3\%$} & \mbox{$+0.6\%$} & \mbox{$-1.9\%$} & \mbox{$-0.2\%$} & \mbox{$+0.4\%$} & \mbox{$-4.5\%$} \\
    
    \mbox{MAE} & \mbox{$-14.3\%$} & \mbox{$+77.1\%$} & \mbox{$-8.6\%$} & \mbox{$-14.3\%$} & \mbox{$+51.4\%$}  & \mbox{$+5.7\%$} & \mbox{$-8.6\%$} & \mbox{$+122.9\%$} \\
    
    \mbox{RMSE} & \mbox{$-13.1\%$}  & \mbox{$+57.0\%$}  & \mbox{$-8.4\%$}  & \mbox{$-12.1\%$}  & \mbox{$+34.6\%$} & \mbox{$+2.8\%$}  & \mbox{$-8.4\%$}  & \mbox{$+85.0\%$} \\
    \bottomrule
    \end{tabular}
    
    \label{tab:2DMBB_accuracy_combinations_percentages}
\end{table}

\newpage

\section{Comparative study for varying sizes of the training dataset}
\label{sec:comparative_data}

In this appendix, we investigate the impact of the number of training points on the achieved accuracy level to provide insights into the selection of the size of the training dataset. To this end, we systematically analyze and compare the performance of our method by varying the size $I$ of the training dataset $\boldsymbol{D}$.

With reference to the 2D MBB beam case study, we randomly sample from the $2500$ reference training instances four additional training sets, consisting of $500$, $1000$, $1500$, and $2000$ data points, respectively. For each dataset, we systematically train new data-driven predictors $\mathcal{NN}_\mathtt{AE}$ and $\mathcal{NN}_\mathtt{FC}$, and assess their predictive capabilities against the same testing data employed for the reference configuration. Below, we report the achieved values of BA, MAE, and RMSE, related to the predicted fields of optimal density (\tab\ref{tab:data_1}), Von Mises stress (\tab\ref{tab:data_2}), and dominant principal stress (\tab\ref{tab:data_3}). For the three performance indicators, two main common trends can be detected across the three prediction scenarios. First, we note a significant improvement in the prediction accuracy when increasing the size of the training dataset from $500$ to $1000$. Second, we observe an accuracy saturation behavior for training sets comprising between $2000$ and $2500$ instances. Accordingly, we have chosen to not increase the size of $\boldsymbol{D}$ beyond $I=2500$, even though the achieved accuracy level is considered satisfactory even for $I=1500$.

\begin{table}[h!]
\small
\caption{MBB beam. Accuracy of the predicted density field for varying sizes of the training dataset against the ground truth SIMP-based procedure. Results are reported in terms of binary accuracy, mean absolute error and root mean squared error, all averaged over 500 testing instances.\label{tab:data_1}}
\centering
  \begin{tabular}{cccccc}
    \toprule
& $I=500$ & $I=1000$ & $I=1500$ & $I=2000$ & $I=2500$ (ref.) \\
\hline
BA & $49.06\%$ & $95.10\%$ & $96.19\%$ & $96.58\%$ & $96.46\%$ \\
MAE & $0.470$ & $0.049$ & $0.038$ & $0.035$ & $0.035$ \\
RMSE & $0.479$ & $0.136$ & $0.112$ & $0.104$ & $0.107$ \\
    \bottomrule
  \end{tabular}
\end{table}

\begin{table}[h!]
\small
\caption{MBB beam. Accuracy of the predicted Von Mises stress field for varying sizes of the training dataset against the ground truth SIMP-based procedure. Results are reported in terms of binary accuracy, mean absolute error and root mean squared error, all averaged over 500 testing instances.\label{tab:data_2}}
\centering
  \begin{tabular}{cccccc}
    \toprule
& $I=500$ & $I=1000$ & $I=1500$ & $I=2000$ & $I=2500$ (ref.) \\
\hline
BA & $99.16\%$ & $99.32\%$ & $99.41\%$ & $99.39\%$ & $99.29\%$ \\
MAE & $0.027$ & $0.022$ & $0.018$ & $0.017$ & $0.018$ \\
RMSE & $0.044$ & $0.036$ & $0.030$ & $0.029$ & $0.030$ \\
    \bottomrule
  \end{tabular}
\end{table}

\begin{table}[h!]
\small
\caption{MBB beam. Accuracy of the predicted dominant principal stress field for varying sizes of the training dataset against the ground truth SIMP-based procedure. Results are reported in terms of binary accuracy, mean absolute error and root mean squared error, all averaged over 500 testing instances.\label{tab:data_3}}
\centering
  \begin{tabular}{cccccc}
    \toprule
& $I=500$ & $I=1000$ & $I=1500$ & $I=2000$ & $I=2500$ (ref.) \\
\hline
BA & $64.97\%$ & $93.98\%$ & $94.91\%$ & $95.35\%$ & $95.42\%$ \\
MAE & $0.032$ & $0.018$ & $0.014$ & $0.013$ & $0.013$ \\
RMSE & $0.042$ & $0.030$ & $0.025$ & $0.024$ & $0.024$ \\
    \bottomrule
  \end{tabular}
\end{table}

\end{document}